\documentclass[12pt]{article}
\usepackage{graphicx,amssymb,amsmath}

\newcommand{\be}{\begin{equation}}
\newcommand{\ee}{\end{equation}}
\newcommand{\bea}{\begin{eqnarray}}
\newcommand{\eea}{\end{eqnarray}}
\newcommand{\beas}{\begin{eqnarray*}}
\newcommand{\eeas}{\end{eqnarray*}}

\begin{document}
\begin{titlepage}

\begin{center}

{\Large Finite $N$ and the failure of bulk locality: Black holes in AdS/CFT}

\vspace{8mm}

\renewcommand\thefootnote{\mbox{$\fnsymbol{footnote}$}}
Daniel Kabat${}^{1}$\footnote{daniel.kabat@lehman.cuny.edu},
Gilad Lifschytz${}^{2}$\footnote{giladl@research.haifa.ac.il}

\vspace{4mm}

${}^1${\small \sl Department of Physics and Astronomy} \\
{\small \sl Lehman College, City University of New York, Bronx NY 10468, USA}

\vspace{2mm}

${}^2${\small \sl Department of Mathematics and Physics} \\
{\small \sl University of Haifa at Oranim, Kiryat Tivon 36006, Israel}

\end{center}

\vspace{8mm}

\noindent
We consider bulk quantum fields in AdS/CFT in the background of an
eternal black hole.  We show that for black holes with finite entropy,
correlation functions of semiclassical bulk operators close to the horizon deviate from their
semiclassical value and are ill-defined inside the horizon.  This is due to the large-time behavior of
correlators in a unitary CFT, and means the region near and inside the
horizon receives corrections.  We give a prescription for modifying
the definition of a bulk field in a black hole background, such that
one can still define operators that mimic the inside of the horizon,
but at the price of violating microcausality.  For supergravity fields we find that commutators
at spacelike separation generically $\sim e^{-S/2}$.  Similar results hold
for stable black holes that form in collapse.  The general lesson may
be that a small amount of non-locality, even over arbitrarily large
spacelike distances, is an essential aspect of non-perturbative
quantum gravity.

\end{titlepage}
\setcounter{footnote}{0}
\renewcommand\thefootnote{\mbox{\arabic{footnote}}}

\section{Introduction\label{sect:intro}}

The holographic principle \cite{'tHooft:1993gx,Susskind:1994vu} states
that in any theory of quantum gravity local bulk physics is only an
illusion.  The physical degrees of freedom can be thought of as living
on a set of codimension-1 hypersurfaces known as holographic screens.
The AdS/CFT correspondence provides a precise realization of this
idea in which the boundary of AdS serves as the holographic screen.
In AdS/CFT the basic claim is that the boundary CFT provides a
complete set of observables, with the CFT Hamiltonian generating the
appropriate unitary time evolution.  Bulk observables, to the extent
that they can be defined, must be expressible in terms of the CFT.\footnote{We assume quantities that cannot be so described are not observables of the bulk theory.}

At least conceptually, a straightforward approach to describing bulk
physics using the CFT is to express bulk quantum fields in terms of
CFT operators.  For free bulk scalar fields the appropriate CFT
operators were constructed in \cite{Balasubramanian:1998sn,Banks:1998dd,
Dobrev:1998md,Bena:1999jv,Hamilton:2005ju,Hamilton:2006az,Hamilton:2006fh},
while for free bulk fields with spin the appropriate CFT operators
were constructed in \cite{Heemskerk:2012mq,Kabat:2012hp}.  These
constructions, which effectively rely on solving free wave equations
in the bulk, can be used to define bulk observables in the leading
large-$N$ limit of the CFT.  The perturbative corrections needed to
take interactions into account were studied in
\cite{Kabat:2011rz,Heemskerk:2012mn} for scalar fields and in
\cite{Kabat:2012av,Kabat:2013wga} for fields with spin.  The corrections were derived using
the $1/N$ expansion of the CFT, which is dual to the perturbative bulk expansion in powers of Newton's constant.

At finite $N$ in the CFT, or equivalently at finite Planck length
in the bulk, it seems clear that any attempt to construct a local bulk
quantum field must fail.  Holographic theories have an entropy bound
\cite{Bousso:1999xy}, and as a result the CFT has far fewer degrees of
freedom than would be necessary to define a local field in the bulk
\cite{Susskind:1998dq}.  Local bulk effective field theory is only an approximation, albeit an 
excellent approximation under ordinary circumstances.

The breakdown of local effective field theory should occur even in a pure AdS background.  For example \cite{Kabat:2011rz,Kabat:2012av,Kabat:2013wga} developed an approach to constructing
interacting bulk fields in the $1/N$ expansion based on enforcing bulk locality.  In these references it was shown that bulk microcausality can be satisfied
to all orders in the $1/N$ expansion.  But microcausality was argued to be violated at finite $N$, even in a pure AdS background, due to effects in the CFT that are non-perturbative in the $1/N$ expansion.
However currently there is no detailed understanding of this.
It might be  easier to understand the failure of the semiclassical approximation in a background where the holographic entropy bound is saturated, most notably, in the background of a black hole.\footnote{Note that the semiclassical approximation must break down when applied to black holes, since the black hole information
paradox cannot be resolved in the context of local effective field theory, see for instance \cite{Mathur:2009hf}.}
This makes the AdS-Schwarzschild geometry a promising arena for exploring the failure of effective field theory.
There are other good motivations for studying this geometry.
Various ideas about the black hole information paradox \cite{Mathur:2005zp,Mathur:2008nj,Almheiri:2012rt,Almheiri:2013hfa} advocate the possibility that the region near or inside the horizon differs from the semiclassical picture, and we would like to understand to what extent
these effects are present in AdS/CFT.

In this paper we use AdS/CFT to motivate the following 
picture of the breakdown of local effective field theory near and inside a black hole horizon: {\em at finite
Planck length, modified continuum bulk quantum fields can still be defined in
terms of the CFT.  Generically these modified fields reproduce semiclassical correlators to a good approximation.
However the modified fields violate microcausality.  That is,
they fail to commute at spacelike separation.\footnote{Bulk gauge
symmetries also lead to commutators which are non-vanishing at spacelike
separation.  This is required by the bulk Gauss constraints
and can be understood from the boundary point of view as arising
from Ward identities in the CFT \cite{Kabat:2012av,Kabat:2013wga}.
But these effects are visible in the $1/N$ expansion and are
perfectly consistent with bulk causality.  By contrast the
finite-$N$ effects we consider in this paper violate bulk
causality.}   Quantities defined in terms of causal structure, such as the
event horizon of a black hole, do not exist at finite Planck length.}

Regarding previous work, non-local effects in quantum gravity have been proposed
by several authors, most notably Giddings \cite{Giddings:2011ks,Giddings:2012gc,Giddings:2013noa},
and mechanisms for the breakdown of bulk locality in AdS/CFT have been studied in \cite{Jevicki:1998rr,Garner:2014kna}.
Many of the results in this paper build on the ideas presented in \cite{Hamilton:2007wj,Lowe:2009mq}.

An outline of this paper is as follows.  In section
\ref{sect:BH} we consider the semiclassical construction of bulk
observables in an AdS-Schwarzschild background.  We point out that the semiclassical construction fails to
give well-defined observables close to and inside the horizon at finite Planck length, and we give a minimal prescription for modifying the
semiclassical construction to obtain observables that are
non-perturbatively well-defined.  In section \ref{sect:Rindler} we study the prescription in more detail in Rindler
coordinates and we give an estimate of the resulting non-perturbative correction
to bulk correlation functions.  We present some explicit calculations for AdS${}_2$ in section \ref{sect:AdS2}
and we comment on BTZ black holes in section \ref{sect:BTZ}.  We conclude in section
\ref{sect:conclusions} by discussing implications of these
results and listing some open questions.  Smearing functions and the bulk geometries we consider are described in appendices
\ref{appendix:BH} and \ref{appendix:coordinates} and some results on CFT correlators are collected in
appendix \ref{appendix:CFT}.

\section{Eternal AdS black holes\label{sect:BH}}

\begin{figure}
\begin{center}
\includegraphics{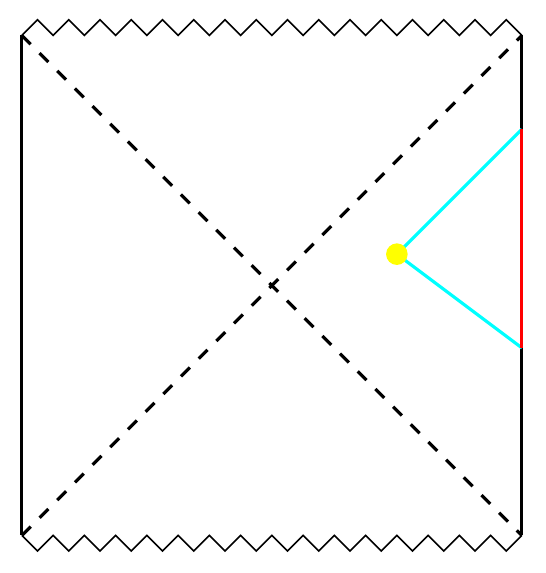}
\end{center}
\caption{An AdS-Schwarzschild black hole with a bulk field operator inserted outside the horizon.  To all orders in $1/N$ the bulk field
can be represented as a sum of CFT operators.  The CFT operators are smeared over a region of the complexified boundary (indicated
in red) which is spacelike separated from the bulk point (indicated in yellow).\label{fig:BHsmear}}
\end{figure}

In this section we study bulk observables in an AdS-Schwarzschild background.
Consider a generic bulk field $\phi$ evaluated at a point outside the horizon of a black hole.  To all orders in the $1/N$
expansion, the bulk field can be expressed as a sum of smeared CFT operators.  We show in appendix
\ref{appendix:BH} that the CFT operators are smeared over a region on the complexified boundary which
is spacelike separated from the bulk point.  This is illustrated in Fig.\ \ref{fig:BHsmear}.

A few comments on our use of complex boundary coordinates are in order.  In empty AdS space one can choose to complexify the boundary spatial coordinates, although one can also
represent fields using data on the real boundary \cite{Hamilton:2006az}.  But in the presence of a black hole one is faced with the problem of reconstructing evanescent waves
\cite{Rey:2014dpa}.  In position space, analytic continuation of the boundary data makes this reconstruction possible.  Alternatively one could remain on the real boundary and frame the discussion in terms of singular distributions as in \cite{Morrison:2014jha}, or one could work in momentum space as in \cite{Papadodimas:2012aq}.  For our
purposes using complex boundary coordinates is convenient, since it makes the discussion below more transparent.

To all orders in $1/N$ the bulk fields constructed in this way obey microcausality.  But at finite $N$, or more precisely when the CFT
has finite entropy, there is an obstruction to implementing microcausality everywhere in the bulk.  The salient observation is that as the bulk point approaches
the future horizon of the black hole, the smearing region extends to future infinity on the boundary.

It's best to think about this in terms of correlation functions.  Consider a bulk-boundary correlator involving one bulk point outside the horizon and some number of
boundary points.  The boundary points are taken to be at arbitrary but finite times.  To all orders in the semiclassical approximation, the bulk-boundary correlator can be
obtained as a sum of smeared CFT correlators.  But as the bulk point approaches the future horizon, the smearing region extends to infinite time on the boundary.\footnote{This effect plays an important role in the computational complexity of \cite{Susskind:2014rva}.}
This means the bulk-boundary correlator becomes sensitive to the late-time behavior of CFT correlators.  At finite entropy this behavior is quite non-trivial and
depends on the details of the CFT spectrum \cite{Maldacena:2001kr,Dyson:2002nt,Barbon:2003aq,Barbon:2014rma}.  In this sense bulk fields near the horizon are fine-grained observables, sensitive to the microstate of the black hole,
and the semiclassical approximation breaks down as one approaches the horizon of the black hole.

\begin{figure}
\begin{center}
\includegraphics[width=14cm]{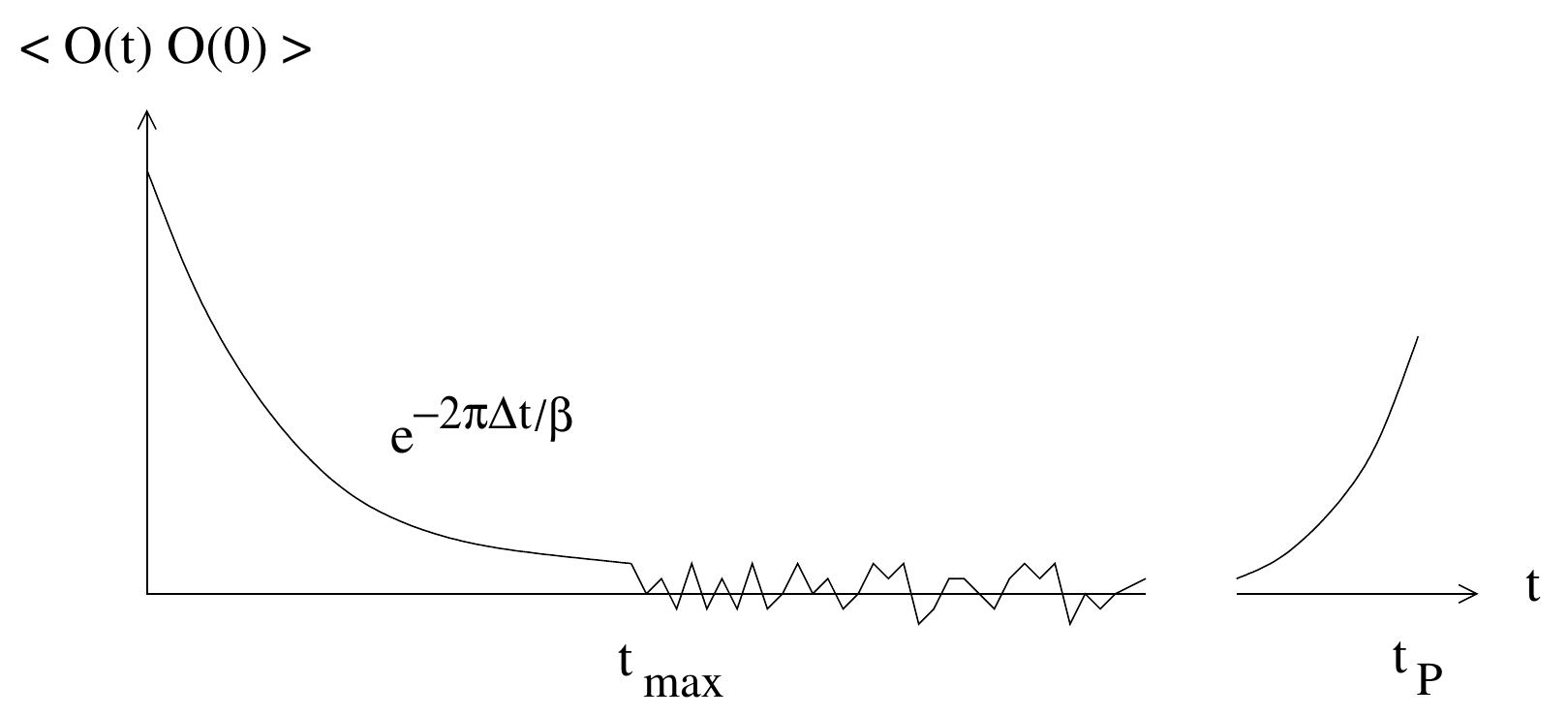}
\end{center}
\caption{Sketch of a CFT correlator in a generic pure state.  The correlator decays exponentially up to $t_{\rm max}$ then begins to fluctuate.
Eventually after a Poincar\'e time the correlator has a large fluctuation.\label{fig:correlator}}
\end{figure}

It's possible to be more precise about the late-time behavior of CFT correlation functions.  In the thermodynamic limit correlators at finite
temperature decay exponentially at late times.  As shown in appendix \ref{appendix:CFT}, for an operator of dimension $\Delta$ the exponential decay is
\be
\label{ExponentialDecay}
\langle {\cal O}(t) {\cal O}(0) \rangle_\beta \sim e^{-2 \pi \Delta \, t / \beta}
\ee
where $\beta$ is the periodicity in imaginary time.  This exponential decay can be thought of as due to excitations dissipating into an infinite heat bath.  But in a system with finite entropy this exponential decay
can't persist forever.  Instead, as pointed out in \cite{Dyson:2002nt}, the correlator can't decay below the generic inner product of two normalized
vectors in the available Hilbert space.\footnote{This statement is corrected by a factor involving the matrix elements of the operators.  We neglect such
multiplicative factors since we're only interested in keeping track of how the result depends on the entropy of the system.}  As discussed in appendix \ref{appendix:timescales}, by picking two unit vectors
at random one finds that on average
\be
\big\vert \langle \psi_1 \vert \psi_2 \rangle \big\vert \sim {1 \over \sqrt{\hbox{\rm dim ${\cal H}$}}} = e^{-S/2}
\ee
where $S$ is the entropy.\footnote{This estimate applies to a correlator in a definite microstate of the CFT, which we assume displays this typical behavior.  It corresponds to the $\vert{\rm noise}\vert_{\rm pure}$ estimate given in (4.9) of \cite{Barbon:2014rma}.}  A more realistic picture of a correlation function is sketched in Fig.\ \ref{fig:correlator}.  The correlator decays exponentially
up to a time $t_{\rm max}$.  After $t_{\rm max}$ the correlator exhibits noisy fluctuations of size set by $e^{-S/2}$.  After a long time, of order the Poincar\'e time $t_{\rm P} \sim \exp(e^S)$, the correlator undergoes a large fluctuation.  The timescale that will be important for us
is $t_{\rm max}$, the time at which correlators stop decaying.

It's easy to estimate $t_{\rm max}$.  By following the exponential decay until the correlator is of order $e^{-S/2}$ we see that
\be
\label{tmax}
t_{\rm max} = {\beta S \over 4 \pi \Delta}
\ee
One can summarize this discussion by saying that after a time $t_{\rm max}$ the system starts to notice that it's living in a finite-dimensional Hilbert space.
After the much longer Heisenberg time $t_{\rm H} \sim \beta e^S$ the system is able to identify its precise microstate.\footnote{The number of states with energy less than $E$
is $n(E) = e^{S(E)}$.  Then ${dn \over dE} = \beta e^S$ and the the spacing between adjacent energy levels is $\Delta E = {1 \over \beta} e^{-S}$.  By the
uncertainty principle, after the time $t_{\rm H} \sim \beta e^S$ one can distinguish individual microstates.}  Finally after the Poincar\'e time
$t_{\rm P} \sim \exp(e^S)$ the correlator undergoes a large fluctuation.

The fact that finite-entropy correlators undergo fluctuations at late times, rather than decaying exponentially, is a problem for defining
bulk observables.  The $1/N$ expansion gives an expression for bulk fields involving integrals over spacelike-separated points on the boundary.
As the bulk point approaches the horizon the region of integration extends to infinite time.  In the $1/N$ expansion this
is acceptable because the entropy diverges and CFT correlators decay.  But at finite $N$ the entropy should be finite. With finite entropy, as the bulk point approaches the horizon the smearing
region will eventually reach $t_{\rm max}$.  At this point bulk correlators will no longer be smooth functions of position.
Instead they will undergo an infinite number of fluctuations as the bulk point approaches the horizon.  Most of these
fluctuations are very small, of order $e^{-S/2}$, but there will also be an infinite number of large fluctuations.  This is certainly not the behavior one would expect from semiclassical reasoning.
Note that this behavior makes the limit as the bulk point approaches the horizon ill-defined.  For bulk points inside the
horizon one has an even worse problem: the semiclassical smearing function grows exponentially with time, see (\ref{InsideRindlerSmear})
for an explicit expression in Rindler coordinates, and when integrated against a fluctuating CFT
correlator one gets completely meaningless expressions. There are exceptions to this rule, for example the Rindler horizons we will study in section \ref{sect:Rindler}.  Rindler horizons have infinite
area and infinite entropy, even at finite $N$, so they do not suffer from this problem and there is no breakdown of the semiclassical approximation near or inside a Rindler horizon.

At this point one could give up and declare that bulk physics near or inside the horizon is not well defined.  However in the holographic approach to quantum gravity one
regards the boundary CFT as primary and thinks of bulk physics as an approximate concept which must be defined in terms of the CFT.  The question
then becomes whether one can give a reasonable prescription for defining bulk observables purely in terms of the CFT.  These bulk observables should be well-defined
close to and perhaps inside the horizon, and they should be reasonable in the sense that they reproduce semiclassical physics up to small corrections.

\begin{figure}
\begin{center}
\hbox{\includegraphics[height = 5cm]{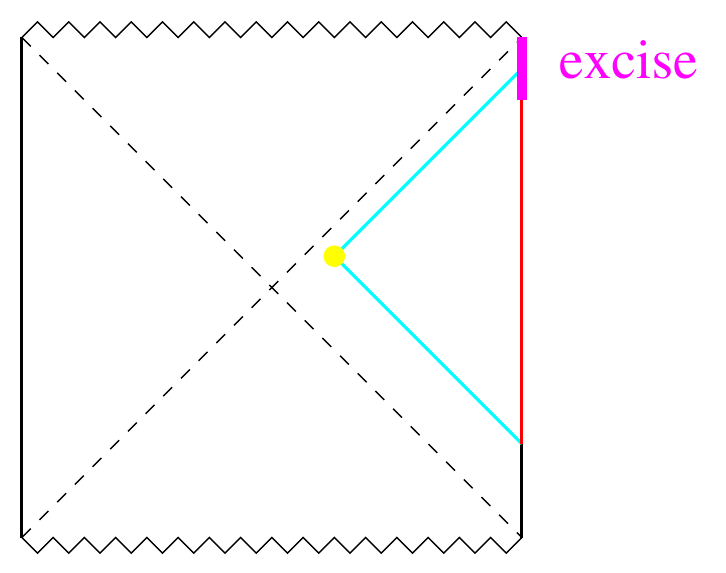} \includegraphics[height=5cm]{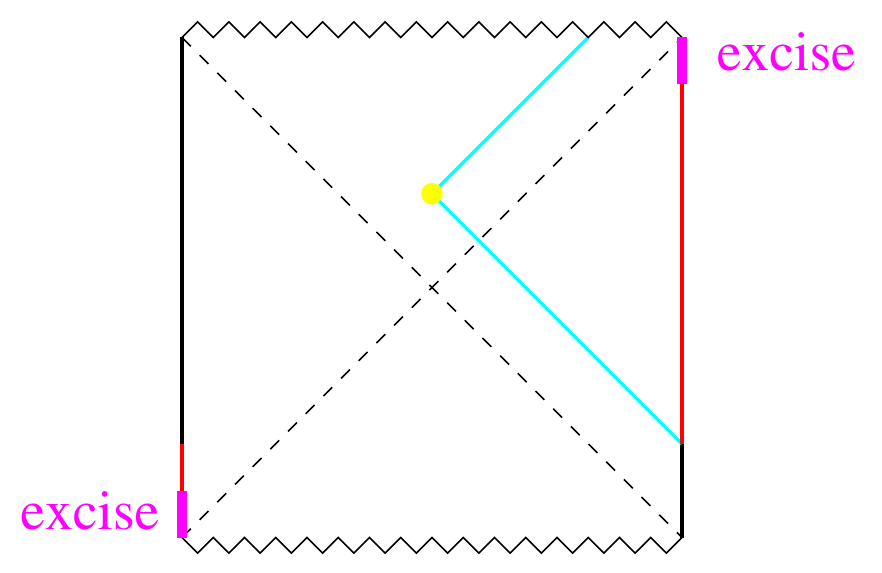}}
\end{center}
\caption{On the left, an AdS-Schwarzschild black hole with a bulk field operator inserted outside the horizon.  The region indicated in purple is excised from the smearing function.  When the bulk point is inside the horizon as on the right, we
expect the smearing function to have support on both boundaries.\label{fig:BHexcise}}
\end{figure}

For an AdS-Schwarzschild black hole there is a reasonable prescription for defining a bulk field which allows us to place an operator near or inside the
horizon. The basic idea was proposed in
\cite{Hamilton:2007wj}.  All we need to do is
excise the late-time region from the smearing function\footnote{There are a few different prescriptions for doing this.  See section \ref{sect:AdS2}.}.  That is, we use the semiclassical expression for a bulk field in terms of the CFT, but by hand we impose a cutoff and
never integrate past $t = t_{\rm max}$ on the boundary.  For bulk points that are inside the horizon we expect that the smearing function will have
support on
both boundaries, and in this case we must excise regions near future infinity on one boundary and near past infinity on the other.  This is illustrated in Fig.\ \ref{fig:BHexcise}.\footnote{Fig.\ \ref{fig:BHexcise} illustrates our expectation for a field of integer conformal dimension, with support at spacelike separation
on the right boundary and timelike separation on the left.  For the case of general dimension inside a Rindler horizon see appendix B of \cite{Hamilton:2006fh}.}

Although it may seem very ad hoc, this prescription has a sensible physical interpretation.  It amounts to modifying the definition of a bulk
operator so as to discard
the part of the CFT correlation function which is sensitive to the detailed microstate structure of the CFT.  In this sense it
corresponds to a ``coarse-graining'' procedure which seems necessary to recover, at least approximately, a well-behaved
notion of bulk physics inside the horizon from the CFT.  Note that this coarse-graining is not just an average over microstates, as is done in some proposals,  rather it is a restriction on the experiments (measurements) one is allowed to perform on the state. Even working with CFT correlators in the canonical ensemble, one has to restrict the allowed experiments in order to obtain an approximate notion of spacetime inside the horizon. One may wonder if each particular microstate in the canonical ensemble is associated with a distinct spacetime geometry in the problematic region. This seems to us very unlikely.  Rather it is much more likely that for individual
microstates, the region close to the semiclassical horizon does not have a geometric (low energy supergravity) description, but requires many additional bulk degrees of freedom.  For example, according to the fuzzball proposal
\cite{Mathur:2005zp,Mathur:2008nj} close to the horizon additional stringy degrees of freedom become important.
The modified smearing functions are not sensitive to these additional degrees of freedom, and this allows them to give an approximate meaning to spacetime inside the horizon.

To say a little more about the cutoff procedure, note that by the uncertainty principle a time cutoff $t_{\rm max}$ corresponds to an energy resolution
\be
\Delta E \sim {1 \over t_{\rm max}} \sim {1 \over \beta S}
\ee
So imposing a time cutoff implies an average over microstates of the CFT with energy differences less than $\Delta E$.
It's interesting to compare this to the energy fluctuations one would expect in, for example, the canonical ensemble.
This is set by the specific heat, which in a CFT is proportional to the entropy.
\be
\Delta E_{\rm canonical} = {1 \over \beta} \sqrt{c} \sim {1 \over \beta} \sqrt{S}
\ee
So the energy resolution allowed by the cutoff procedure is much finer than the fluctuations present in the canonical ensemble.
The key feature of the cutoff prescription is not so much that it enforces an average over microstates.  Rather it
discards the late-time behavior of the CFT correlator, which seems necessary to recover bulk physics, and which
a mere average over microstates does not seem to do.  For example in the canonical ensemble, or equivalently in the thermofield double state of the CFT,
correlation functions display qualitatively similar noisy behavior at late times \cite{Dyson:2002nt,Barbon:2003aq,Barbon:2014rma}.

By definition, this prescription makes bulk observables well-defined.  It remains to show that the prescription is reasonable, in the sense of giving small corrections to semiclassical expectations.
Ideally we'd show this for AdS-Schwarzschild.  But for computational simplicity, in the next section we will instead study this for the simpler but analogous case of pure AdS in Rindler coordinates.  Even at this stage, however,
there are a few preliminary remarks worth making.
\begin{itemize}
\item
According to this prescription, correlators are only modified when the bulk point gets sufficiently close to the horizon.  This fits with the idea
that there is some non-trivial structure at the horizon.  But unlike the firewall proposal \cite{Almheiri:2012rt,Almheiri:2013hfa}, the modification
to correlators at the horizon is quite mild.\footnote{This lack of drama at the horizon is not surprising, since we are studying an eternal black hole
(dual to a thermofield double state in the CFT) which is not expected to have a firewall.  To study firewalls one would have to consider
more generic entangled states in the doubled CFT Hilbert space \cite{Marolf:2013dba}.}  In this sense our proposal is more inline with the fuzzball philosophy \cite{Mathur:2011wg}.
\item
The region that is excised is timelike-separated from all points on the AdS boundary.  This means that, at least generically, bulk operators near or inside the horizon will not commute
with any local operators in the CFT: a clear violation of bulk micro-causality.
\item
The fact that bulk operators inside the horizon do not commute with operators on the boundary can be interpreted as saying that global horizons do not exist at finite Planck length.
That is, in asymptotically flat space one defines the horizon as the boundary of the causal past of ${\cal I}^+$, or equivalently as the boundary of the region where local fields commute with all
operators on ${\cal I}^+$.  With asymptotic AdS boundary conditions the horizon is the boundary of the causal past of the timelike AdS boundary, and operators inside the horizon should commute with operators on the boundary at
late times.  With our prescription, such a region does not exist when the CFT has finite entropy.\footnote{One can
reach the same conclusion from the following point of view. In the presence of a horizon the CFT seems to have quasinormal modes, signaling complex poles in a CFT two-point function \cite{Birmingham:2002ph}. However a CFT with finite entropy must have discrete energy levels, thus any two-point function should only have poles on the real axis.}
\end{itemize}

\section{Rindler coordinates\label{sect:Rindler}}

In this section we study the excision procedure and the resulting change in correlators in the simpler setting of AdS
in Rindler coordinates.  The coordinates we use are presented in appendix \ref{appendix:coordinates}.

To be clear about our motivation, note that a Rindler horizon has infinite area.  So the CFT has infinite entropy even at finite $N$, and the discussion in section \ref{sect:BH} about the late-time behavior of CFT correlators doesn't apply.  In fact Rindler CFT correlators decay exponentially even at late times, so there is no real need to modify the Rindler smearing functions.  This is all consistent with the fact that nothing special happens
at a Rindler horizon.  Our motivation in this section is not to use Rindler coordinates to study the late-time behavior of CFT correlators.  Rather we are using them to ask: suppose we excise a late-time region from the
smearing functions.  How does this affect bulk correlators?

\begin{figure}
\begin{center}
\includegraphics[height=6cm]{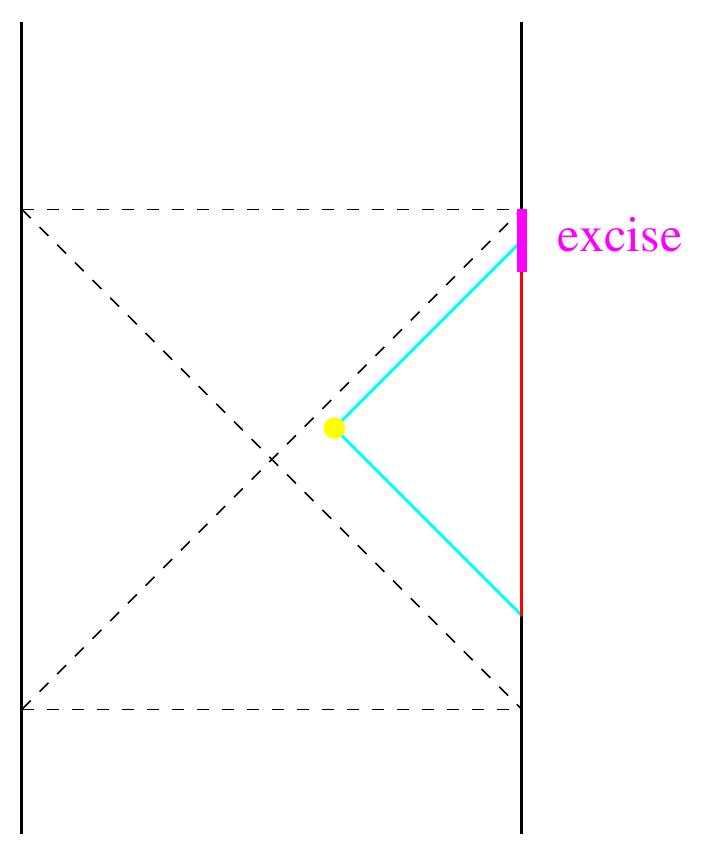} \hspace{12mm} \includegraphics[height=6cm]{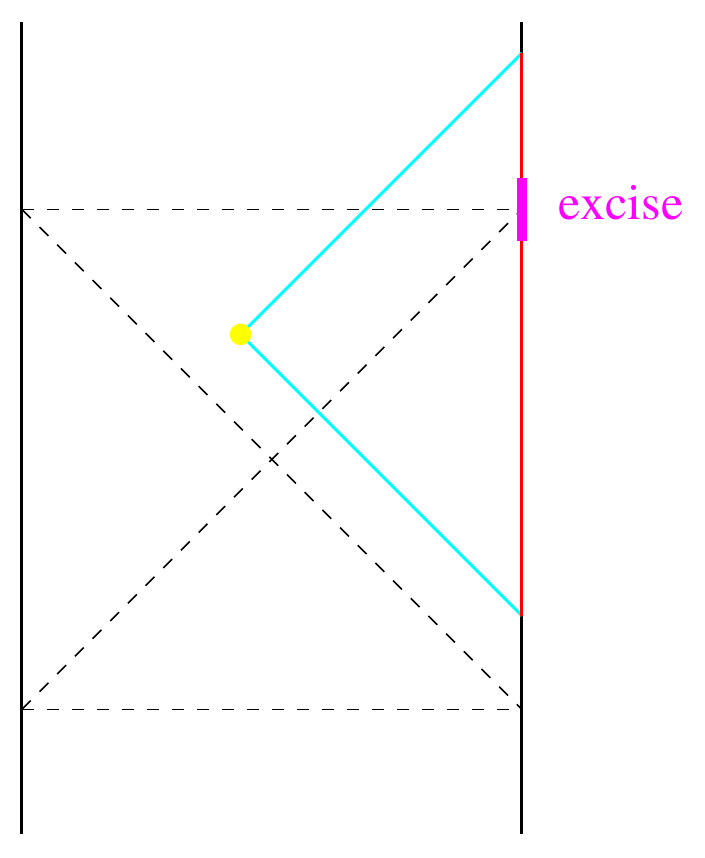}
\end{center}
\caption{A slice of AdS in Rindler coordinates, drawn as an AdS${}_2$ Penrose diagram.  The diagonal dashed lines are the Rindler horizons at $r = R$ and the horizontal dashed lines
are the would-be singularities at $r = 0$.  On the left, an operator
just outside the Rindler horizon.  On the right, an operator inside the Rindler horizon.\label{fig:RindlerExcise}}
\end{figure}

In Rindler coordinates the metric on AdS${}_{d+1}$ reads
\bea
\label{AdS-Rindler}
&& ds^2 = - {r^2 - R^2 \over R^2} dt^2 + {R^2 \over r^2 - R^2} dr^2 + r^2 \left(d\phi^2 + \sinh^2 \phi \, d\Omega_{d-2}^2\right) \\[3pt]
\nonumber
&& -\infty < t < \infty \qquad R < r < \infty \qquad 0 < \phi < \infty
\eea
The quantity in parenthesis is the metric on the hyperbolic plane ${\cal H}^{d-1}$,
\be
ds^2_{{\cal H}^{d-1}} = d\phi^2 + \sinh^2 \phi \, d\Omega_{d-2}^2
\ee
To obtain a smearing function in this geometry it's convenient to analytically continue $\phi = i \theta$.  Under this continuation
\be
ds^2_{{\cal H}^{d-1}} \rightarrow - d\theta^2 - \sin^2 \theta \, d\Omega_{d-2}^2 = - d\Omega^2_{d-1}
\ee
where we're taking $0 < \theta < \pi$.  That is, aside from an overall change of sign of the metric, the continuation turns ${\cal H}^{d-1}$ into $S^{d-1}$.
The AdS metric becomes
\be
ds^2 =  - {r^2 - R^2 \over R^2} dt^2 + {R^2 \over r^2 - R^2} dr^2 - r^2 d\Omega^2_{d-1}
\ee
This is de Sitter space in static coordinates.\footnote{The static patch is $0 < r < R$, where $t$ is timelike and the metric is static.}  The timelike boundary of AdS
becomes the past boundary of de Sitter space.  Up to a divergent conformal factor the induced metric on the past boundary is $ds^2 = dt^2 + R^2 d\Omega^2_{d-1}$.
That is, the boundary is ${\mathbb R} \times S^{d-1}$ which can be conformally compactified to $S^d$.
The field at a bulk point outside the AdS horizon, meaning at $r > R$, can be expressed in terms of data on the past de Sitter boundary using a retarded Green's function.
From the AdS point of view this means bulk fields outside the horizon can be expressed using a smearing function with support at spacelike separation on the complexified
boundary.  This is indicated in the left panel of Fig.\ \ref{fig:RindlerExcise}.  Note that for bulk points outside the horizon, the smearing function can cover at most half of the
past de Sitter boundary, namely the region\footnote{To see this one starts at $r = R$ and sends a null geodesic to the past in the $\theta$ direction.  When the geodesic reaches $r = \infty$ it has covered a range $\Delta \theta = \pi/2$.}
\be
\label{MaxSupport}
-\infty < t < \infty \qquad 0 < \theta < \pi/2
\ee

We'll also want to consider bulk points inside the horizon.  Nothing special happens at a Rindler horizon, so there can be no difficulty in representing a field at $r < R$.
To make this manifest it's useful to switch to Poincar\'e coordinates, since these coordinates cover a larger patch of AdS.  In Poincar\'e coordinates the AdS metric is
\be
ds^2 = {R^2 \over Z^2} \left(-dT^2 + \vert d{\bf X} \vert^2 + dZ^2\right) \qquad 0 < Z < \infty
\ee
To represent bulk fields in Poincar\'e coordinates we continue ${\bf X} = i {\bf Y}$, which turns the AdS metric into
\be
ds^2 = {R^2 \over Z^2} \left(-dT^2 - \vert d{\bf Y} \vert^2 + dZ^2\right) \qquad 0 < Z < \infty
\ee
This is de Sitter space in planar or inflationary coordinates, with $Z$ playing the role of conformal time.  The past de Sitter boundary is at $Z = 0$, with induced metric (up to a divergent
conformal factor) $ds^2 = dT^2 + \vert d{\bf Y} \vert^2$.  In other words the past boundary of de Sitter is ${\mathbb R}^d$ which again can be conformally compactified to $S^d$.  A field in the bulk can
be expressed in terms of data on the past boundary using a retarded Green's function.  In AdS this corresponds to spacelike separation on the complexified boundary.  But if the bulk
point is inside the Rindler horizon, the smearing region will extend past the Rindler patch of the boundary.\footnote{We'll
use Poincar\'e coordinates on the boundary so this will cause no difficulty.  If one insists on using Rindler coordinates
one can use the antipodal map (\ref{antipodal}) to move the part of the smearing function that extends outside the Rindler patch over
to the left Rindler boundary \cite{Hamilton:2005ju,Hamilton:2006fh}.}  This is shown in the right panel of Fig.\ \ref{fig:RindlerExcise}.

\begin{figure}
\begin{center}
\includegraphics{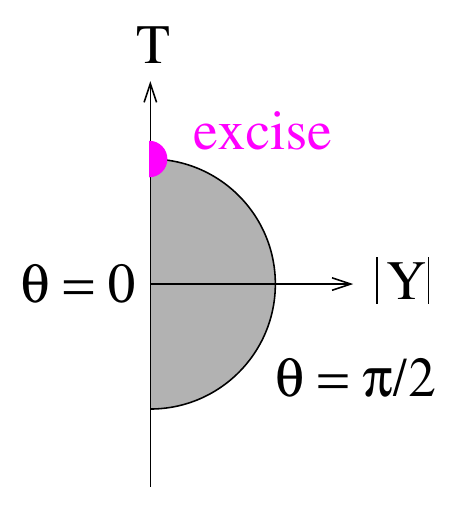}
\end{center}
\caption{The region of the complexified boundary needed to describe bulk points outside the Rindler horizon is shown in gray.  In Poincar\'e
coordinates it's a ball of radius $R$, $T^2 + \vert {\bf Y} \vert^2 \leq R^2$.  The region that gets excised is is indicated in magenta.
It's a tiny ball about the point $T = R$, ${\bf Y} = 0$.\label{fig:region}}
\end{figure}

We'll need the relation between Rindler and Poincar\'e coordinates on the complexified boundary.  After analytic
continuation, it follows from (\ref{t}) and (\ref{phi}) in appendix \ref{appendix:coordinates} that
\bea
\label{RPt}
&& \tanh(t/R) = {2 R T \over R^2 + T^2 + \vert {\bf Y} \vert^2} \\
\label{RPtheta}
&& \tan \theta \, {\bf n} = {2 R {\bf Y} \over R^2 - T^2 - \vert {\bf Y} \vert^2}
\eea
Here ${\bf n} \in S^{d-2}$, $\vert {\bf n} \vert = 1$.  Only part of the past de Sitter boundary is visible from points in
de Sitter space with $r > R$.  This region was described in (\ref{MaxSupport}), namely
\be
-\infty < t < \infty \qquad 0 < \theta < \pi/2
\ee
This defines the largest region where a smearing function can have support
for bulk points that are outside the Rindler horizon.
Switching to Poincar\'e coordinates, this region is a ball of radius $R$, given by $T^2 + \vert {\bf Y} \vert^2 \leq R^2$.  This is shown in Fig.\ \ref{fig:region}.\footnote{It helps to note that from (\ref{RPtheta}) curves of fixed $\theta$ are circles in the $(T,\vert{\bf Y}\vert)$ plane,
centered at $(T = 0,\vert{\bf Y}\vert = - R / \tan \theta)$ and with radius $R / \sin \theta$.  These circles all pass through the points
$(T = \pm R,\vert{\bf Y}\vert = 0)$.}

Now we can study the excision procedure.  By the usual Euclidean continuation the AdS-Rindler geometry (\ref{AdS-Rindler}) is thermal with inverse
temperature $\beta = 2 \pi R$.  If we adopt the estimate (\ref{tmax}), we'd say we should impose a late-time cutoff on the smearing functions at\footnote{This is not the only possible prescription for introducing a cutoff, but it's adequate for our purposes here.  A few other possible prescriptions are discussed in section \ref{sect:AdS2}.}
\be
t_{\rm max} = {\beta S \over 4 \pi \Delta} = {R S \over 2 \Delta}
\ee
Note that $S$ here does not refer to the entropy of the Rindler horizon, since the Rindler entropy is infinite.  Rather we're introducing $S$ as a convenient way to parametrize the cutoff.

The first question we ask is, how close can we get to the Rindler horizon before the cutoff starts to matter?  To study this consider a bulk operator inserted
at a point $(t,r)$ outside the horizon.  By following light rays to the boundary we find that the smearing function extends to the future of
$t$ by an amount
\be
\delta t = {1 \over 2} R \log {r + R \over r - R}
\ee
Given a cutoff at $\delta t = t_{\rm max}$, this means we can probe the region
\be
r > {R \over \tanh (t_{\rm max}/R)}
\ee
before worrying about the cutoff.  In terms of the entropy this means
\be
r \gtrsim R \left(1 + 2 e^{-S/\Delta}\right)\,.
\ee
So we can go exponentially close to the horizon before the cutoff makes any difference.

Once the bulk point is very close to or inside the horizon, how are correlation functions affected?  To study this consider the correlator between
one bulk point and an arbitrary number of boundary points.  We excise the region
$t > t_{\rm max}$ from the smearing function for the bulk operator.  Rindler coordinates have a coordinate singularity at $t = \infty$,
so to study the effect of the excision it's convenient to switch to Poincar\'e coordinates.
Expanding (\ref{RPt}) about $t = \infty$, the excised region $t > t_{\rm max}$ corresponds to a ball
\be
(T-R)^2 + \vert {\bf Y} \vert^2 < 4 R^2 e^{-2 t_{\rm max} / R}
\ee
on the complexified Poincar\'e boundary.  This is shown in Fig.\ \ref{fig:region}.  In terms of entropy the excised ball has radius $2 R e^{-S/2\Delta}$.
To understand what this excision means, we consider two cases in turn.

\begin{figure}
\begin{center}
\includegraphics[height=6cm]{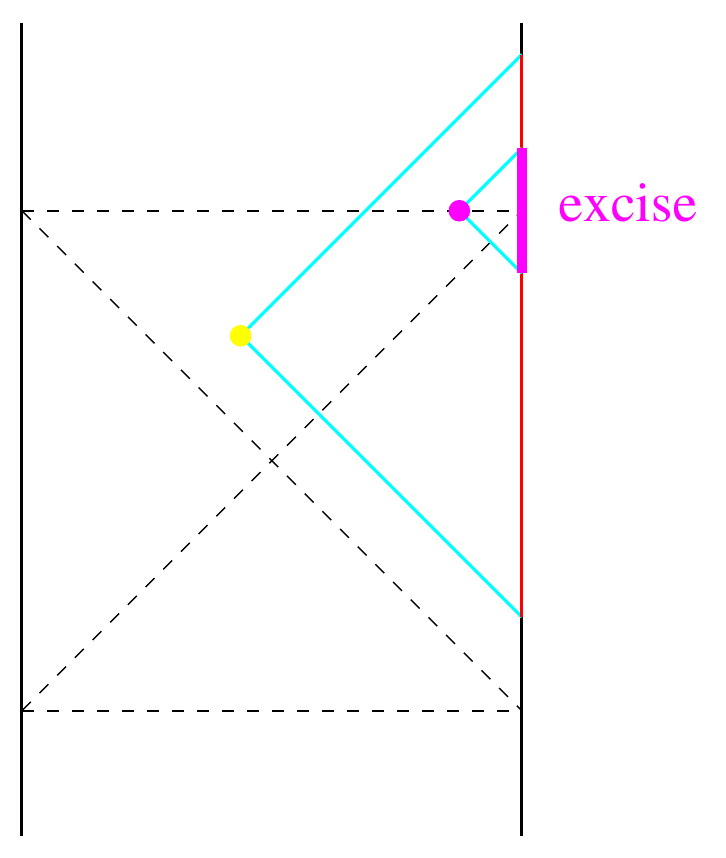}
\end{center}
\caption{For free massless fields the excised region on the boundary is dual to a local operator in the bulk.
The excised bulk operator, shown in magenta, is inserted near $r = 0$.\label{fig:MasslessExcise}}
\end{figure}

\noindent{\em Massless fields} \\
Consider a field in the bulk dual to an
operator ${\cal O}$ with dimension $\Delta = d$.  Examples of such fields are free massless scalars and linearized metric perturbations.
These are simple cases to consider because when $\Delta = d$ the smearing function is constant.\footnote{More precisely it's a step function,
zero at timelike separation and constant at spacelike separation.  See (\ref{PoincareSmear2}).}  Suppose the
bulk point is inserted deep enough inside the horizon that the smearing region completely overlaps with the excised region,
as in the right panel of Fig.\ \ref{fig:RindlerExcise}.  Then what's being excised is the integral of ${\cal O}$ over a ball of radius
$2 R e^{-S/2\Delta}$ on the Poincar\'e boundary.  But for $\Delta = d$, this is exactly the CFT representation of a local field in the
bulk!  In Poincar\'e coordinates the excision corresponds to a bulk operator located at $T = R$, ${\bf X} = 0$ with a radial position set by the radius of the excised region, namely
\be
\label{Z}
Z = 2 R e^{-S/2\Delta}\,.
\ee
Translating this into Rindler coordinates using appendix \ref{appendix:coordinates},
one finds that the excised bulk operator is located at
\be
t = {RS \over 2\Delta} \qquad r = R e^{-S/2\Delta} \qquad \phi = 0
\ee
So the excised bulk operator is inside the Rindler horizon, very close to the would-be Rindler singularity at $r = 0$ and also close to
the right Rindler boundary (since inside the horizon $t \rightarrow \infty$ on the right boundary).  This is illustrated in Fig.\ \ref{fig:MasslessExcise}.  To state this excision in terms of the
CFT we recall that in Poincar\'e coordinates $\phi \sim Z^\Delta {\cal O}$ as $Z \rightarrow 0$.  So to a good approximation what's
being excised is a local operator in Poincar\'e coordinates.  Since $Z \sim e^{-S/2\Delta}$, the excised operator is proportional to
\be
\label{massless}
e^{-S/2} {\cal O}(T = R, {\bf X} = 0)
\ee

\noindent{\em General case} \\
When $\Delta \not= d$ the smearing function is not constant, and in general the smearing region may not completely overlap with the excised region on the boundary.  Moreover once interactions are turned on a bulk field corresponds to a tower of higher-dimension smeared operators in the CFT.  So in
general the excision does not have a simple interpretation as a local operator in the bulk.  Instead the excision
corresponds to a complicated superposition of bulk fields inserted at small $r$ and large $t$.  It's simpler to think about the excision
in terms of the CFT.  Working in Poincar\'e coordinates, for large entropy we see from Fig.\ \ref{fig:region} that
to a very good approximation the excised region can be represented as a local operator
inserted at $T = R$, ${\bf X} = 0$ (the point on the boundary which corresponds to $t \rightarrow \infty$ in Rindler coordinates).
Roughly speaking we've modified the definition of a bulk field for points close to or inside the Rindler horizon,
\be
\label{modified}
\phi_{\rm modified} = \phi_{\rm semiclassical} - e^{-dS/2\Delta} {\cal O}(T = R,{\bf X} = 0)
\ee
where again ${\cal O}$ is a local operator on the Poincar\'e boundary, and the prefactor $e^{-dS/2\Delta}$ comes from the volume of the
excised region.  Note that ${\cal O}$ in general doesn't have a well-defined scaling dimension.  This result reduces to (\ref{massless}) in
the special case $\Delta = d$.

Now it's easy to understand how the excision procedure affects correlation functions.  By construction the modified bulk field
$\phi_{\rm modified}$ is insensitive to the late-time behavior of CFT correlators.  So to evaluate correlators involving
(\ref{modified}) we're free to make up whatever late-time
behavior we like.  It's convenient to pretend that CFT correlators behave semiclassically and decay exponentially at late times.  In Rindler space
there's no need to pretend, since CFT correlators really do decay exponentially at late times.

First consider the correlator between a bulk point close to or inside
the horizon and an arbitrary number of boundary points.  The boundary points are taken to be at fixed finite times.  From (\ref{modified}) the excision procedure changes the correlator by $e^{-dS/2\Delta}$ times a correlator
in the CFT involving ${\cal O}$ which we take to be ${\cal O}(1)$.  Operators of large dimension are more sensitive to the excision.  But it's particularly interesting to
consider massless supergravity fields which are dual to operators of dimension $\Delta = d$.  For these fields\footnote{For example a massless scalar field is dual to an operator with $\Delta = d$, and the bulk graviton is dual to the CFT stress tensor which also
has $\Delta = d$.  A special case would seem to be bulk gauge fields which are dual to conserved currents of dimension $d - 1$.  However the smearing function
for gauge fields has support on a spherical shell $S^{d-1}$ (the intersection of the past lightcone with the de Sitter boundary) rather than on a ball $B^d$ \cite{Kabat:2012hp}.  For such a smearing function the excised volume $\sim ({\rm radius})^{d-1}$ and the estimate of the change in a correlator is again given by (\ref{change}).} the change in correlators is generically of
order
\be
\label{change}
e^{-S/2}
\ee
Note that the region we're excising is timelike separated from all points on the Rindler boundary.  This means generically the operator ${\cal O}$ we're subtracting
in (\ref{modified}) will not commute with any local operator on the Rindler boundary.  So for bulk points close to or inside the horizon, $\phi_{\rm modified}$
will not commute with local operators on the boundary.  Generically the commutator will be non-zero and (for massless supergravity fields)
of order $e^{-S/2}$.  This is a dramatic breakdown
of bulk microcausality.  Note that the breakdown extends all the way out to the AdS boundary, which is at infinite spacelike separation from points in the bulk!

Since we have in mind an AdS-Schwarzschild black hole with an entropy that is ${\cal O}(N^2)$ , the change in bulk correlators we have found is tiny.  Generically the correction is proportional to
\be
e^{-dS/2\Delta} \sim e^{-{\rm const.} N^2}
\ee
This would correspond to a non-perturbative effect in the $1/N$ expansion, or equivalently a non-perturbative effect
in bulk quantum gravity.  But the $1/N$ expansion
respects bulk locality.  So correction we have identified, although tiny, would appear to be the leading non-perturbative
effect which spoils bulk locality.

It's important to note, however, that the correction is not always small.  In particular consider the correlator between a bulk field close to or inside the horizon, and a local operator on the boundary which is inserted at very late Rindler times.  The boundary operator can have coincident or lightcone singularities with the operator ${\cal O}$ we're subtracting in (\ref{modified}), and this singularity can overcome
the $e^{-dS/2\Delta}$ suppression.  So the change in correlators relative to semiclassical expectations can be arbitrarily large, when an operator is inserted at very late Rindler times on the boundary.

This additional singularity provides
another way to see that global horizons do not exist at finite Planck length.  A local operator outside the horizon would
have two lightcone singularities with operators on the boundary: one where the past lightcone of the bulk point touches the boundary, and another where the future lightcone touches the boundary.  Semiclassically, for an operator inside the horizon, only the past lightcone can touch the boundary.  But we have just argued that, given the
modified bulk operators, a second singularity is indeed present at finite $N$.  This can be interpreted as saying there
is no horizon at finite $N$.

\section{Calculations in AdS${}_2$\label{sect:AdS2}}

As a simple calculable example we consider the case of AdS${}_2$ in Rindler coordinates, with CFT operators of dimension $\Delta=1$.
For notational simplicity, in this section we set the AdS radius of curvature $R = 1$.

We start with the CFT two-point function
\begin{equation}
\langle{\cal O}_R(t){\cal O}_R(t')\rangle=\frac{1}{2(1-\cosh(t-t'))}
\end{equation}
This is relevant for two operators on the right boundary in the thermofield double (TFD) formalism.  An operator ${\cal O}_L$ on the left boundary
can be obtained by shifting $t \rightarrow t + i \pi$.  Thus for two operators on different boundaries
\begin{equation}
\langle{\cal O}_{L}(t){\cal O}_{R}(t')\rangle=\frac{1}{2(1+\cosh(t-t'))}
\end{equation}
Note that we're taking time to run upward on the right boundary and downward on the left boundary.

\subsection{Bulk point outside the horizon}

For a bulk point in the right Rindler wedge
\be
\phi(t,r) = {1 \over 2} \int_{t - \delta t}^{t + \delta t} dt' {\cal O}_R(t')
\ee
where the range of the smearing is set by
\be
\label{Rrange}
\delta t = {1 \over 2} \log {r + 1 \over r - 1}
\ee
This means the bulk-boundary two-point function is
\bea
\nonumber
\langle \phi(t,r) {\cal O}_R(t^{\prime\prime}) \rangle & = &
{1 \over 4} \int_{t - \delta t}^{t + \delta t} dt' \, {1 \over 1 - \cosh(t'-t^{\prime\prime})} \\
\nonumber
& = & {1 \over 4} \coth \left({t + \delta t - t^{\prime\prime} \over 2}\right) -
{1 \over 4} \coth \left({t - \delta t - t^{\prime\prime} \over 2}\right) \\
& = & {1 \over 2 \left(r - \sqrt{r^2 - 1} \, \cosh(t - t^{\prime\prime})\right)}
\eea
where we used $\cosh \delta t = {r \over \sqrt{r^2 - 1}}$.  Likewise for a bulk point in the right Rindler wedge
and a local operator on the left boundary
\be
\langle \phi(t,r) {\cal O}_L(t^{\prime\prime}) \rangle = {1 \over 2 \left(r + \sqrt{r^2 - 1} \, \cosh(t - t^{\prime\prime})\right)}
\ee

Since we're interested in probing the future horizon let's put a cutoff on the smearing integral at $t = t_{\rm cut}$.
There are various prescriptions one could use to fix the cutoff time.  One could set $t_{\rm cut} = t_{\rm max}$,
where $t_{\rm max}$ is defined in (\ref{tmax}).  Another possibility is to set $t_{\rm cut} = t^{\prime\prime} + t_{\rm max}$,
since that's a better estimate of when the correlator starts to become noisy, but then the definition of the bulk operator depends on the
position of the other operator.  A third possibility is to restrict the range of
the smearing integral by setting $t_{\rm cut} = t - \delta t + t_{\rm max}$, which may have advantages for black holes as discussed
below (\ref{texrex}).

\begin{figure}
\begin{center}
\includegraphics[height=6cm]{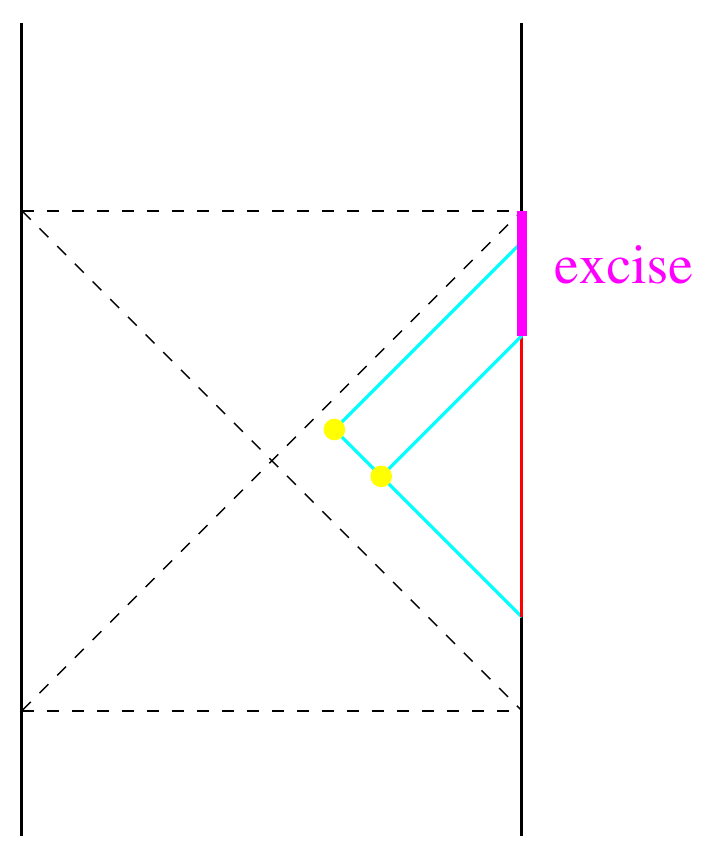} \hspace{5mm} \includegraphics[height=6cm]{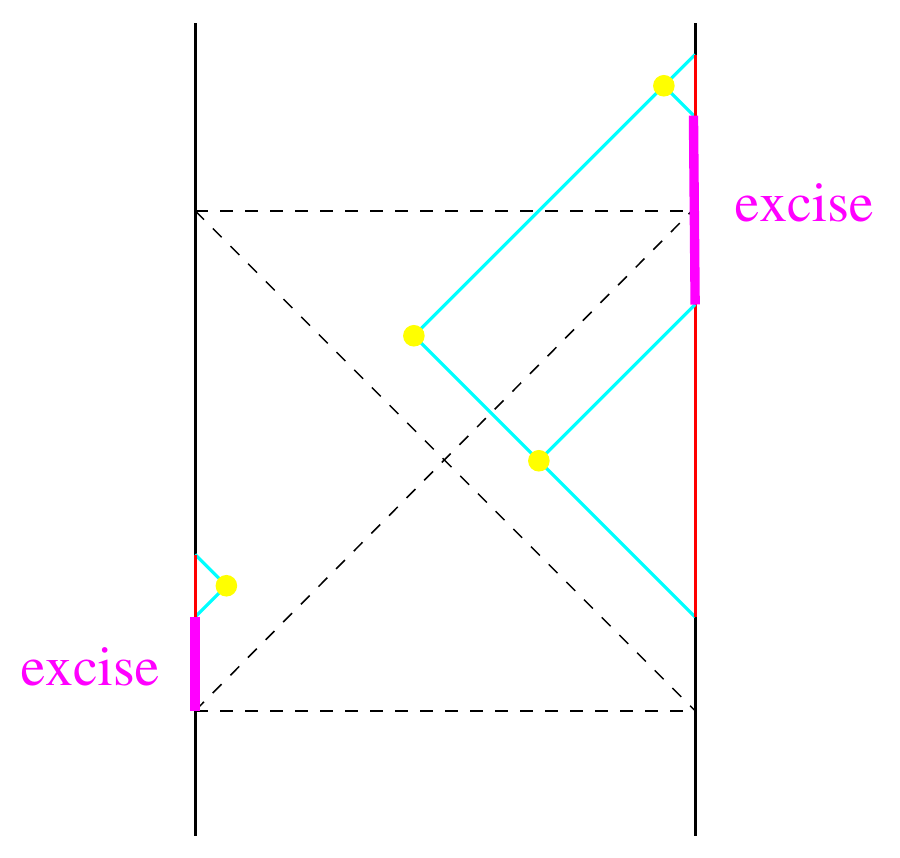}
\end{center}
\caption{For a bulk point in the right Rindler wedge the modified bulk operator is a local field inserted at a new position.
For a bulk point in the future wedge the modified operator is a superposition of a local operator on the right and a local
operator on the left.\label{fig:AdS2}}
\end{figure}

With the cutoff in place we have the modified correlator
\bea
\langle \phi(t,r) {\cal O}_R(t^{\prime\prime}) \rangle & = &
{1 \over 4} \int_{t - \delta t}^{t_{\rm cut}} dt' \, {1 \over 1 - \cosh(t'-t^{\prime\prime})} \\
\nonumber
& = & {1 \over 4} \coth \left({t_{\rm cut} - t^{\prime\prime} \over 2}\right) -
{1 \over 4} \coth \left({t - \delta t - t^{\prime\prime} \over 2}\right)
\eea
As can be seen in the left panel of Fig.\ \ref{fig:AdS2}, this is the semiclassical result one would obtain for a
bulk operator inserted at a new position
\be
t_{\rm new} = {t_{\rm cut} + t - \delta t \over 2} \qquad r_{\rm new} = \coth {t_{\rm cut} - (t - \delta t) \over 2}
\ee
Note that the modified correlator has singularities at
\be
\label{Rsing}
t^{\prime\prime} = t_{\rm cut} \qquad {\rm and} \qquad t^{\prime\prime} = t - \delta t
\ee
This is very special for this example. In general the modified smearing does not lead to an expression that looks like a bulk point with a modified position, although the fact that the position of the singularities is modified is generally correct.
One can also think of this procedure as defining a modified bulk field
\be
\phi_{\rm modified}(t,r) = \phi_{\rm semiclass}(t,r) - \phi_{\rm semiclass}(t_{\rm ex},r_{\rm ex})
\ee
where we're excising a bulk operator located at
\be
t_{\rm ex} = {t + \delta t + t_{\rm cut} \over 2} \qquad r_{\rm ex} = \coth {t + \delta t - t_{\rm cut} \over 2}
\ee
This means the change in the correlator due to the cutoff is
\be
\langle \phi_{\rm semiclass}(t_{\rm ex},r_{\rm ex}) {\cal O}_R(t^{\prime\prime})\rangle =
{1 \over 2 \left(r_{\rm ex} - \sqrt{r_{\rm ex}^2 - 1} \cosh(t_{\rm ex} - t^{\prime\prime})\right)}
\ee
This is generically small as long as $t_{\rm ex} \gg t^{\prime\prime}$.  Again this is special for the case treated here ($\Delta=1$ in AdS${}_2$). In
general the correction does not look like it is coming from an extra local operator but the position of the new singularity and the size of the correction are similar.

\subsection{Bulk point inside the horizon}

Now consider a bulk point in the future Rindler wedge.  For $\Delta = 1$ the smearing is given by
\be
\label{InsideSmear}
\phi(t,r) = {1 \over 2} \int_{t - \delta t}^\infty dt' {\cal O}_R(t') - {1 \over 2} \int_{t + \delta t}^\infty dt'
{\cal O}_L(t')
\ee
where instead of (\ref{Rrange}) the range of smearing is set by
\be
\delta t = {1 \over 2} \log {1 + r \over 1 - r}
\ee
Note that time runs upward on the right boundary and downward on the left.  The smearing is over spacelike
separated points on the right boundary and timelike separated points on the left.  The relative $(-)$ sign in
(\ref{InsideSmear}) comes from the factor $(-1)^\Delta$ associated with the antipodal map that was used to move
part of the smearing over to the left boundary.

Semiclassically this representation for a bulk field leads to the bulk-boundary correlator
\bea
\nonumber
\langle \phi(t,r) {\cal O}_R(t^{\prime\prime})\rangle & = &
{1 \over 2} \int_{t - \delta t}^\infty dt' \langle {\cal O}_R(t') {\cal O}_R(t'') \rangle
- {1 \over 2} \int_{t + \delta t}^\infty dt' \langle {\cal O}_L(t') {\cal O}_R(t'') \rangle \\
& = & {1 \over 2 \left(r - \sqrt{1 - r^2} \sinh(t - t^{\prime\prime})\right)}
\eea
We define a modified bulk field by introducing cutoffs on the left and right boundaries.
\be
\label{InsideSmearmod}
\phi_{\rm modified}(t,r) \equiv {1 \over 2} \int_{t - \delta t}^{t_{\rm cut,\,R}} dt' {\cal O}_R(t') - {1 \over 2} \int_{t + \delta t}^{t_{\rm cut,\,L}} dt'
{\cal O}_L(t')
\ee
As can be seen in the right panel of Fig.\ \ref{fig:AdS2}, the modified field is equivalent to a pair of bulk operators,
one inserted in the right Rindler wedge and the other inserted in the left wedge.  This makes it clear that the modified
correlator with an operator on the right boundary has singularities at
\be
\label{InsideSing}
t^{\prime\prime} = t_{\rm cut,\,R} \qquad {\rm and} \qquad t^{\prime\prime} = t - \delta t
\ee
The modification can also be thought of as defining
\be
\phi_{\rm modified}(t,r) = \phi_{\rm semiclass}(t,r) - \phi_{\rm semiclass}(t_{\rm ex},r_{\rm ex})
\label{modifiedinside}
\ee
where we're excising a bulk operator inserted in the future wedge at
\be
\label{texrex}
t_{\rm ex} = {t_{\rm cut,\,L} + t_{\rm cut,\,R} \over 2} \qquad
r_{\rm ex} = \tanh {t_{\rm cut,\,L} - t_{\rm cut,\,R} \over 2}
\ee
With the prescription of equal cutoffs on the left and right boundaries the excised operator is at $r_{\rm ex} = 0$.
Alternatively with the prescription $t_{\rm cut,\,R} = t - \delta t + t_{\rm max}$, $t_{\rm cut,\,L} = t + \delta t + t_{\rm max}$
the excised operator is at the same value of $r$ as the original operator, $r_{\rm ex} = r$.  This prescription may be advantageous for
black holes, since it avoids placing the excised operator at the singularity.  In any case the modified
smearing makes a correction to the correlator given by
\be
\langle \phi_{\rm semiclass}(t_{\rm ex},r_{\rm ex}) {\cal O}_R(t^{\prime\prime})\rangle =
{1 \over 2 \left(r_{\rm ex} - \sqrt{1 - r_{\rm ex}^2} \sinh(t_{\rm ex} - t^{\prime\prime})\right)}
\ee
This is generically small as long as $t_{\rm ex} \gg t^{\prime\prime}$.  The size of the modification, generically
$e^{-t_{\rm ex}} \sim e^{-t_{\rm max}}$, agrees with the general estimate (\ref{change}).

These results show that for a large class of operators one can get a reasonable approximation to the spacetime
near or inside the horizon.  However if the boundary operator itself is inserted at very late times there are additional singularities at finite $t^{\prime\prime}$, given in (\ref{Rsing}), (\ref{InsideSing}), which are not present in the semiclassical result.  This means the causal structure has changed, and indeed there is no event horizon. To see this recall that one property of the event horizon is that an operator on or inside the future horizon has a singularity with an
operator on the right boundary only where the past lightcone of the bulk point hits the boundary.
However with the modified smearing there are two times when the bulk-boundary correlator is singular.  This means
there is no event horizon, since operators on or inside the would-be event horizon do not commute with boundary operators at late (but finite) times.

\subsection{Two bulk points inside the horizon}

If one tries to compute a correlator with two modified bulk operators inserted inside the horizon using the prescription (\ref{InsideSmearmod}), the result can differ significantly from what one expects semiclassically. This can be seen from (\ref{modifiedinside}) by noting that the contribution to the correlation function from the two operators $\phi_{\rm semiclass}(t_{\rm ex},r_{\rm ex})$ can be large. 
A simple but not entirely satisfactory way to circumvent this problem is to choose different $t_{\rm cut}$ prescriptions for each of the bulk operators. Then the correlation function of the extra operators will be small if the values of $t_{\rm cut}$ are different enough. 

Another way to avoid the problem is to note that there are two distinct ways to modify the bulk operator inside the horizon.  This is because there are two equivalent ways of representing a semiclassical bulk operator inside the horizon. One way is to keep the smearing on the right boundary and use the antipodal map to shift the rest to the left, which gives the representation (\ref{InsideSmear}). The other way is to keep the smearing on the left boundary and move the rest to the right using the antipodal map. This results in an alternate (but equivalent) representation of a bulk operator inside the horizon,
\be
\label{InsideSmearalt}
\phi^{\rm alt}(t,r) = {1 \over 2} \int_{-\infty}^{t +\delta t} dt' {\cal O}_L(t') - {1 \over 2} \int^{t - \delta t}_{-\infty }dt'
{\cal O}_R(t')
\ee
Now if we define a modified operator by cutting off the smearing region,
\be
\label{InsideSmearaltmod}
\phi^{\rm alt}_{\rm modified}(t,r) = {1 \over 2} \int_{-t_{\rm cut,\,L}}^{t +\delta t} dt' {\cal O}_L(t') - {1 \over 2} \int^{t - \delta t}_{-t_{\rm cut,\,R} }dt'
{\cal O}_R(t')
\ee
then a two-point function of two modified operators, one with representation (\ref{InsideSmearaltmod}) and one with representation (\ref{InsideSmearmod}), will deviate only slightly from the semiclassical result.  This approach does have the drawback that the representation
of one bulk operator depends on the representation chosen for the other operator in the correlator.

\section{Comments on BTZ\label{sect:BTZ}}

In this section we extend the discussion to AdS black holes with hyperbolic horizons,
following the earlier work \cite{Hamilton:2007wj,Lowe:2009mq}.

In the Rindler coordinates of section \ref{sect:Rindler} and appendix \ref{appendix:coordinates}
the AdS metric is
\be
ds^2 = - {r^2 - R^2 \over R^2} dt^2 + {R^2 \over r^2 - R^2} dr^2 + r^2 ds^2_{{\cal H}^{d-1}}
\ee
The Rindler horizon at $r = R$ is a non-compact hyperbolic space ${\cal H}^{d-1}$.
Rindler horizons are just coordinate artifacts.  But one can quotient by a freely-acting subgroup of the
$SO(d-1,1)$ isometries of ${\cal H}^{d-1}$ to make an AdS black hole whose horizon is a compact hyperbolic
manifold \cite{Mann:1997iz,Birmingham:1998nr}.  These are genuine black holes, in which the CFT
lives in finite volume and has finite entropy once $N$ is finite.  Much of our discussion can be carried through without
modification and applies to this case.  Here we make a few remarks on the extension.

As a simple prototype example we consider the BTZ black hole.  It's conventional to rescale the coordinates,
setting\footnote{In \cite{Hamilton:2007wj} hats were used to denote a different set of rescaled coordinates.  Sorry.}
\be
t = r_0 \hat{t} / R \qquad r = R \hat{r} / r_0 \qquad \phi = r_0 \hat{\phi} / R
\ee
where $r_0$ is an arbitrary parameter with units of length.  This puts the AdS metric in the form
\be
ds^2 = - {\hat{r}^2 - r_0^2 \over R^2} d\hat{t}^2 + {R^2 \over \hat{r}^2 - r_0^2} d\hat{r}^2 + \hat{r}^2 d\hat{\phi}^2
\ee
Periodically identifying $\hat{\phi} \approx \hat{\phi} + 2\pi$, or $\phi \approx \phi + 2 \pi r_0/R$, gives a BTZ black hole with a horizon at $\hat{r} = r_0$.

Bulk observables in this geometry were considered in \cite{Hamilton:2007wj} and \cite{Lowe:2009mq}, and this
section is largely a summary of previous results.  It's quite straightforward to construct bulk observables
because in the $1/N$ expansion we can use the same smearing functions as in AdS.  To see this,
note that if a smearing function is integrated against a boundary correlator that has the correct $2\pi$ periodicity
in $\hat{\phi}$, it will automatically produce a bulk correlator that also has the correct periodicity.  So there's no need
to change the smearing functions.

To fix ideas we review the steps to recover a free bulk-to-boundary correlator from the CFT \cite{Hamilton:2006fh}.
Consider the correlator between a bulk point $(t,r,\phi)$ that is inside the horizon and a point on the right boundary at $(t',\phi')$.  For simplicity
we set $t = \phi = 0$.  Applying the smearing function (\ref{InsideRindlerSmear}) to the CFT correlators (\ref{Rindler2point}), (\ref{Rindler2pointLR}) allows us to recover the bulk-to-boundary correlator
in AdS${}_3$, given by \cite{Ichinose:1994rg}
\bea
\nonumber
\langle \phi(t = \phi = 0,r) {\cal O}(t',\phi') \rangle_{\rm AdS} &=& \lim_{r' \rightarrow \infty} \left({r' \over R^2}\right)^\Delta {1 \over 4 \pi R \sqrt{\sigma^2 - 1}}\,
{1 \over \left(\sigma + \sqrt{\sigma^2 - 1}\,\right)^{\Delta - 1}} \\
&\sim& \left(r \cosh \phi' + \sqrt{R^2 - r^2} \sinh (t'/R)\right)^{-\Delta}
\eea
Then we use the fact that in the semiclassical limit, correlators in the BTZ geometry can be represented as an image sum \cite{Lifschytz:1993eb}.  For example
the BTZ bulk -- boundary correlator can be written as
\be
\label{ImageSum}
\langle \phi(t,r,\phi) {\cal O}(t',\phi') \rangle_{\rm BTZ} = \sum_{n = - \infty}^\infty
\langle \phi(t,r,\phi) {\cal O}(t',\phi' + 2 \pi n r_0 / R) \rangle_{\rm AdS}
\ee
The AdS correlator decays exponentially at large $\phi'$, so the image sum is nicely convergent.\footnote{Note that to get convergent expressions
for bulk points inside the horizon, one should first perform the smearing integral then do the image sum.}

Now let's study the effect of imposing a cutoff on the smearing functions at $t = t_{\rm max}$.  To do this, consider representing the right hand side
of (\ref{ImageSum}) in terms of CFT correlators.  The CFT correlator (\ref{Rindler2point})
\be
\langle {\cal O}(t,\phi) {\cal O}(t',\phi') \rangle \sim \left(\cosh(\phi - \phi') - \cosh{t - t' \over R} \right)^{-\Delta}
\ee
decays exponentially when $\vert \phi - \phi' \vert > \vert t - t' \vert / R$.  So imposing a cutoff on the smearing functions at $t = t_{\rm max}$
is like imposing a cutoff on the image sum at $\vert \phi' + 2 \pi n r_0 / R \vert \approx t_{\rm max}/R$.  Away from
the BTZ singularity the image sum is exponentially convergent, so the additional cutoff only makes an exponentially
small effect.  But the BTZ singularity is a fixed point of the identification $\phi \approx \phi + 2\pi r_0 / R$, and
semiclassically the image sum diverges as $r \rightarrow 0$.  So near $r = 0$ the additional cutoff has a large effect,
that it eliminates the $r \rightarrow 0$ singularity in correlation functions!  This was studied in \cite{Hamilton:2007wj},
where it was found that the cutoff only becomes important at a radius
\be
r \approx R e^{-t_{\rm max}/R}
\ee
which is exponentially close to the singularity.

\section{Implications for bulk physics\label{sect:conclusions}}

We have shown that at finite entropy the late-time behavior of CFT
correlators is an obstruction to defining local quantum fields in the
bulk.  Local bulk fields can be defined to all orders in the $1/N$
expansion, but the semiclassical representation of bulk fields in
terms of CFT operators leads to ill-defined correlators near and
inside the horizon of a black hole that has finite entropy.  We gave a
minimal prescription for modifying the definition of a bulk field in
order to get correlators which are well-defined near or inside the
horizon.  The prescription we adopted, of imposing a cutoff on the
smearing functions at late times, discards the part of the boundary
correlator which is sensitive to the microstates of the CFT. This
leads to well-defined correlators, but there is a price that must be
paid.  There are small deviations from semiclassical correlators as a
bulk point approaches the horizon, and these deviations imply a
failure of bulk locality: the modified bulk operators fail to commute
at spacelike separation, by an amount that is generically of order
$e^{-S/2}$ for massless supergravity fields.

Our results leave many open questions.  We gave a minimal prescription
for modifying the definition of a bulk field which allowed us to probe
the horizon of an AdS-Schwarzschild black hole.  But is there any
sense in which the prescription is unique or preferred?  Also the
prescription allowed us to consider correlators involving one or two bulk
points near or inside the horizon and an arbitrary number of boundary
points.  Is there a prescription that gives well-defined correlators
involving any number of points inside the horizon?  Finally it is expected
that even in a pure AdS background microcausality will be violated due
to finite $N$ effects in the CFT.  Is this
violation related to the effects studied in the present
paper?

Given our results, an immediate consequence is that there are no
global horizons at finite Planck length.  In classical gravity one
considers the causal past of the AdS boundary and defines the horizon
as the boundary of this region.  Equivalently the horizon is the
boundary of the region where all local operators commute with all
operators on the AdS boundary at late times.  If we take this
definition over into the quantum theory, we have just shown that there
is no such region at finite Planck length.

Let us see what we can conclude about black holes formed by
collapse. We can construct semiclassical bulk operators using the
$1/N$ expansion that are appropriate for a collapsing black hole
background.  When used in a finite-$N$ CFT these semiclassical bulk
operators will reproduce semiclassical results to a good approximation
as long as the range of smearing on the boundary is not too large,
i.e.\ as long as one does not probe too close to the horizon. If one
is very close to the horizon then correlation functions of these bulk
operators will start deviating from the semiclassical result.  This
shows that there is some structure near the horizon, similar to the
fuzzball idea.  Once correlation functions deviate from their
semiclassical form, we expect that microcausality (as defined by the
semiclassical black hole background) will break down.
The small features in CFT correlators at late times encode which
particular microstate one is in, thus they encode unitary time
evolution. One could capture this information using the semiclassical
bulk operators as long as we are outside the horizon (though the
resulting correlators will differ from the semiclassical result). 
For bulk points inside the horizon the semiclassical smearing functions
grow exponentially on the boundary at late times.  For a black hole formed in collapse
one can use the mirror operators of \cite{Papadodimas:2012aq,Papadodimas:2013jku} to
construct the ${\cal O}_{L}$.  But because of the exponential growth of the smearing kernel,
we see that information about the CFT
microstate, as encoded in CFT correlators, obstructs the existence of
bulk operators inside the horizon of a black hole formed in collapse.
One then might say that the region inside the
horizon does not exist.  This is similar to the conclusion reached in
\cite{Almheiri:2012rt,Almheiri:2013hfa}.
This, however, is not the end of our story. We saw that we could
define modified bulk operators which throw away the late-time behavior
of CFT correlators.  This allowed us to define bulk operators inside
the horizon of an eternal black hole, and this is obviously also possible for black holes formed by collapse.  
This makes sense since if there is an inside of the
horizon it must be independent of the particular microstate. We
achieved this not by averaging over microstates, but by using
operators which are insensitive to the particular microstate one
considers.

Whether we use the modified operators or the original semiclassical
ones, given the sensitivity of bulk microcausality to the details of the CFT correlators, there will be some breakdown of bulk locality near the horizon.
It would be interesting to understand this better.  There are
important conceptual questions to address, such as whether it is
possible to build a sensible  bulk theory that violates
causality.\footnote{The CFT is a well-behaved quantum system, so the
question is just about finding a consistent interpretation of the
bulk.}  It presumably helps that the violations of causality are
tiny, generically of order $e^{-S/2}$ for massless supergravity
fields.\footnote{However the causality violations are not always
small.  As pointed out in section \ref{sect:Rindler}, operators
inserted on the boundary at late times can have arbitrarily large
commutators with operators near or inside the black hole horizon.}
It would be interesting to understand if the causality violation
expected in empty AdS is related to the more robust effects studied in
this paper.  It could be the two effects are the same, but that it's
easier to characterize the causality violation in a background like a
black hole in which the holographic bound is saturated.

Finally it would be interesting to study the implications of our
results for the puzzles surrounding black hole evaporation
\cite{Hawking:1976ra} and firewalls
\cite{Almheiri:2012rt,Almheiri:2013hfa}. To study evaporating black holes one has to overcome two obstacles. The first obstacle is that we need to know
the smearing functions appropriate to an evaporating black hole. The second is that that an evaporating black hole in AdS is dual to a non-typical state in the CFT, and thus we have only a limited understanding of its properties.
Nevertheless, although we only treated stable AdS-Schwarzschild black holes (including black holes formed in collapse), it's tempting to
speculate that our results are more general than their derivation, and
that even for evaporating black holes in asymptotically flat space one
will find non-vanishing commutators at spacelike separation.  This would represent a breakdown of bulk locality, as discussed in
\cite{Lowe:2006xm,Papadodimas:2013kwa}, and would imply a new type of uncertainty principle, where a local
measurement far from the black hole could disturb the black hole
interior.  Assuming the commutator is of order $e^{-S/2}$, a local
measurement would disturb the interior by an amount $\sim e^{-S/2}$.
Following Page \cite{Page:1993df,Page:1993wv} one must observe at
least half the Hawking radiation to get any information about the
black hole interior.  This entails at least $e^{+S/2}$ measurements,
and by the above uncertainty principle this would seem to make an
${\cal O}(1)$ disturbance to the black hole interior.  This may be
connected to the ideas of black hole complementarity
\cite{Susskind:1993if,Susskind:1993mu}.  It's also curious that
non-local commutators of order $e^{-S/2}$ seem capable of accounting
for the pairwise correlations between outgoing Hawking particles that
must be present in order for black hole evaporation to be a unitary
process \cite{Iizuka:2013ria}.

\bigskip
\bigskip
\goodbreak
\centerline{\bf Acknowledgements}
\noindent
We are grateful to Tom Banks, Lam Hui, Janna Levin, Don Marolf, Joe Polchinski
and Vladimir Rosenhaus for valuable discussions and to Nori Iizuka for detailed comments
on the manuscript.  DK is supported by U.S.\
National Science Foundation grant PHY-1125915 and by grants from
PSC-CUNY.  The work of GL was supported in part by the Israel Science
Foundation under grants 392/09 and 504/13 and in part by a grant from
GIF, the German-Israeli Foundation for Scientific Research and
Development under grant 1156-124.7/2011.

\appendix
\section{Smearing in AdS-Schwarzschild\label{appendix:BH}}

In this appendix we consider the problem of representing a bulk quantum field in an AdS-Schwarzschild
geometry in terms of the CFT.  Our goal is to show that, to all orders in $1/N$, the bulk field
can be represented as a sum of CFT operators which are smeared over a spacelike-separated region on the complexified
boundary.

We'll consider fields in the AdS-Schwarzschild geometry \cite{Hawking:1982dh,Witten:1998zw}
\bea
&& ds^2 = -f dt^2 + {1 \over f} dr^2 + r^2 d\Omega^2_{S^{d-1}} \\
\nonumber
&& f(r) = {r^2 \over R^2} + 1 - {\omega_d M \over r^{d-2}}
\eea
Here $d\Omega^2_{S^{d-1}}$ is the metric on a round unit $(d-1)$-sphere, which we write as
\be
d\Omega^2_{S^{d-1}} = d\theta^2 + \sin^2 \theta \, d\Omega^2_{S^{d-2}} \qquad 0 < \theta < \pi
\ee
Also $M$ is the black hole mass, $\omega_d = {16 \pi G_N \over (d-1) {\rm vol}(S^{d-1})}$, and $R$ is the AdS
radius of curvature.  The black hole horizon is located at $r = r_0$, where $f(r_0) = 0$.

To get started, suppose the bulk field obeys a free wave equation.  We'd like to express the field at a point $(t,r,\theta)$ outside the horizon
in terms of data on the right asymptotic boundary.  Without loss of generality we take $\theta = 0$.  To obtain an expression for the bulk field
it's convenient to follow \cite{Hamilton:2006az,Hamilton:2006fh} and analytically continue the
spatial coordinates, setting $\theta = -i \phi$.  Under this continuation
\be
d\Omega^2_{S^{d-1}} \rightarrow - d\phi^2 - \sinh^2 \phi \, d\Omega^2_{S^{d-2}} \qquad 0 < \phi < \infty
\ee
Aside from a change of sign, this is the metric on hyperbolic space ${\cal H}^{d-1}$.  This means the AdS-Schwarzschild geometry continues to
\be
ds^2 = - f dt^2 + {1 \over f} dr^2 - r^2 ds^2_{{\cal H}^{d-1}}
\ee
This continued geometry is somewhat curious.  Since the $(t,r)$ part of the metric hasn't been changed the Penrose diagram looks like the diagram for an AdS-Schwarzschild
black hole (Fig.\ \ref{fig:BHsmear}), but rotated 90${}^\circ$, and with an ${\cal H}^{d-1}$ fiber over each point.
Outside the horizon $r$ plays the role of a time coordinate, and the boundary
at $r \rightarrow \infty$ becomes the past boundary of de Sitter space.\footnote{When $M = 0$ the geometry is just
de Sitter in an unconventional slicing.  To see this introduce coordinates on the de Sitter hyperboloid
$-u^2 - v^2 - \vert {\bf x} \vert^2 + y^2 = -R^2$ by setting $u = \sqrt{r^2 + R^2} \cos(t/R)$, $v = \sqrt{r^2 + R^2}
\sin(t/R)$, ${\bf x} = r \sinh \phi \, {\bf n}$, $y = r \cosh \phi$.}

The bulk field can be expressed in terms of data on the past de Sitter boundary using a retarded
Green's function.  Of course the field only depends on data in the past lightcone of the bulk point.  Returning to anti-de Sitter space, this means the bulk field outside the horizon
can be expressed in terms of data at spacelike separation on the complexified boundary.  One gets an expression of the form
\be
\label{ComplexSmear}
\phi(t,r,\theta = 0) = \int_{\rm spacelike} dt' d\phi' d\Omega'_{d-2} \, K(t,r,\theta = 0 \vert t',\phi',\Omega') {\cal O}(t',\phi',\Omega')
\ee
Here $\Omega' \in S^{d-2}$, and the integral is over points on the complexified boundary that are spacelike separated from the bulk point.

To take interactions into account we follow \cite{Kabat:2011rz} and imagine adding an infinite tower of higher-dimension multi-trace operators to the definition of the bulk field.
\be
\phi = \int K {\cal O} + \sum_i a_i \int K_i {\cal O}_i
\ee
Here $K_i$ is the smearing function appropriate to the operator ${\cal O}_i$.  Order-by-order in the $1/N$ expansion the coefficients of the higher-dimension
operators $a_i$ can be chosen to obtain bulk fields that commute at spacelike separation.  This gives the desired result, that to all orders in $1/N$ the bulk field
can be represented as a sum of CFT operators smeared over a spacelike-separated region on the complexified
boundary.

\section{Smearing in Rindler and Poincar\'e coordinates\label{appendix:coordinates}}

In this appendix we describe AdS using Rindler and Poincar\'e coordinates and we collect some results on smearing
functions.  For a related discussion of AdS in Rindler coordinates see \cite{Parikh:2012kg}.

AdS${}_{d+1}$ is a hypersurface in ${\mathbb R}^{2,d}$ defined by
\be
-u^2 - v^2 + \vert {\bf x} \vert^2 + y^2 = - R^2
\ee
To describe this in Rindler or accelerating coordinates we set
\bea
&& u = r \cosh \phi \\
&& {\bf x} = r \sinh \phi \, {\bf n} \\
\label{v}
&& v = \sqrt{r^2 - R^2} \sinh(t/R) \\
\label{y}
&& y = \sqrt{r^2 - R^2} \cosh(t/R)
\eea
where ${\bf n} \in S^{d-2}$, $\vert {\bf n} \vert = 1$.  The induced metric is
\be
\label{RindlerMetric}
ds^2 = - {r^2 - R^2 \over R^2} dt^2 + {R^2 \over r^2 - R^2} dr^2 + r^2 d\phi^2 + r^2 \sinh^2 \phi \, d\Omega_{d-2}^2
\ee
Here $-\infty < t < \infty$, $R < r < \infty$, $0 < \phi < \infty$.  We'll also be interested in Poincar\'e coordinates, defined by
\bea
&& u = {R^2 + Z^2 - T^2 + \vert {\bf X} \vert^2 \over 2 Z} \\
&& v = {R T \over Z} \\
&& {\bf x} = {R {\bf X} \over Z} \\
&& y = {R^2 - Z^2 + T^2 - \vert {\bf X} \vert^2 \over 2 Z}
\eea
for which the induced metric is
\be
\label{PoincareMetric}
ds^2 = {R^2 \over Z^2} \left(-dT^2 + \vert d{\bf X} \vert^2 + dZ^2\right)
\ee
with $0 < Z < \infty$.  On the boundary $r \rightarrow \infty$, $Z \rightarrow 0$ and it follows that these coordinates are related by
\bea
\label{t}
&& \tanh(t/R) = {2 R T \over R^2 + T^2 - \vert {\bf X} \vert^2} \\
\label{phi}
&& \tanh \phi \, {\bf n} = {2 R {\bf X} \over R^2 - T^2 + \vert {\bf X} \vert^2}
\eea

It's useful to introduce the AdS-invariant distance $\sigma$ between two points, which we define in the embedding
space by
\be
\sigma = {1 \over 2 R^2} \left[-(u-u')^2 - (v-v')^2 + \vert {\bf x} - {\bf x}' \vert^2 + (y - y')^2\right] + 1
\ee
In Poincar\'e coordinates
\be
\sigma = {1 \over 2 Z Z'} \big(Z^2 + Z'{}^2 + \vert {\bf X} - {\bf X}' \vert^2 - (T - T')^2 \big)
\ee
while in Rindler coordinates for two points outside the horizon
\be
\sigma = {r r' \over R^2} \left(\cosh \phi \cosh \phi' - \sinh \phi \sinh \phi' \, {\bf n} \cdot {\bf n'}\right)
- {1 \over R^2} \sqrt{(r^2 - R^2)(r'{}^2 - R^2)} \, \cosh {t - t' \over R}
\ee
To obtain the distance for points inside the future horizon we modify (\ref{v}), (\ref{y}) slightly and define
\bea
&& v = \sqrt{R^2 - r^2} \cosh(t/R) \\
&& y = \sqrt{R^2 - r^2} \sinh(t/R)
\eea
for $0 < r < R$.  Then the invariant distance between a point $(t,r,\phi,{\bf n})$ inside the future horizon and
a point $(t',r',\phi',{\bf n}')$ in the right Rindler wedge is
\be
\label{FRsigma}
\sigma = {r r' \over R^2} \left(\cosh \phi \cosh \phi' - \sinh \phi \sinh \phi' \, {\bf n} \cdot {\bf n'}\right)
- {1 \over R^2} \sqrt{(R^2 - r^2)(r'{}^2 - R^2)} \, \sinh {t - t' \over R}
\ee
Note that $\sigma$ grows exponentially as $t' \rightarrow \infty$.

This AdS-invariant distance is useful because the smearing functions can be expressed quite simply
in terms of $\sigma$.  For example in Poincar\'e coordinates a free field in the bulk is represented as
\be
\label{PoincareSmear}
\phi(T,{\bf X},Z) = \int\limits_{\rm spacelike} dT' d^{d-1} Y' \, K {\cal O}
\ee
where the integral is over points at spacelike separation on a slice of the complexified boundary.
The boundary metric on this slice is $ds^2 = dT^2 + \vert d{\bf Y} \vert^2$ and the smearing function is
\cite{Hamilton:2006fh}
\be
K = c_{d\Delta} \lim_{Z' \rightarrow 0} (\sigma Z')^{\Delta - d}
\ee
The normalization
\be
\label{cdDelta}
c_{d\Delta} = {2^{\Delta - d} \Gamma(\Delta - {d \over 2} + 1) \over \pi^{d/2} \Gamma(\Delta - d + 1)}
\ee
is fixed so that
\be
\phi(T,{\bf X},Z) \sim Z^\Delta {\cal O}(T,{\bf X}) \qquad \hbox{\rm as $Z \rightarrow 0$}
\ee
Just to write (\ref{PoincareSmear}) completely explicitly,
\be
\label{PoincareSmear2}
\phi(T,{\bf X},Z) = {\Gamma(\Delta - {d \over 2} + 1) \over \pi^{d/2} \Gamma(\Delta - d + 1)}
\int\limits_{T'{}^2 + \vert {\bf Y}' \vert^2 < Z^2} \!\!\!\!\!\!\! dT' d^{d-1} Y'
\left({Z^2 - T'{}^2- \vert {\bf Y}' \vert^2 \over Z}\right)^{\Delta - d}
{\cal O}(T + T',{\bf X} + i {\bf Y}')
\ee

In Rindler coordinates we use the normalization
\be
\label{RindlerNormalization}
\phi(t,r,\phi,{\bf n}) \sim \left({R^2 \over r}\right)^\Delta {\cal O}(t,\phi,{\bf n}) \qquad \hbox{\rm as $r \rightarrow \infty$}
\ee
A free bulk field is represented by
\be
\phi(t,r,\phi,{\bf n}) = \int\limits_{\rm spacelike} dt' d\Omega'_{d-1} R^{d-1} \, K {\cal O}
\ee
where the integral is over points at spacelike separation on a slice of the complexified Rindler boundary.
The boundary metric on this slice is $ds^2 = dt^2 + R^2 d\Omega^2_{d-1}$ and the smearing function is
\be
\label{RindlerSmear}
K = c_{d\Delta} \lim_{r' \rightarrow \infty} \left({R^2 \sigma \over r'}\right)^{\Delta - d}
\ee
To write an explicit expression it's convenient to first use the manifest ${\mathbb R} \times SO(1,d-1)$ isometry
present in Rindler coordinates to place the bulk point at $t = \phi = 0$ with ${\bf n}$ arbitrary.
Then for a point outside the horizon one has
\bea
&& \phi(t=\phi=0,r > R) \\
\nonumber
&& = c_{d\Delta}
\int_{\rm spacelike} dt' d\theta' d\Omega'_{d-2} R^{d-1} \sin^{d-2} \theta' \left(r \cos \theta' - \sqrt{r^2 - R^2} \cosh {t' \over R}\right)^{\Delta-d}
{\cal O}(t',i\theta',{\bf n}')
\eea
where the spacelike region on the boundary is characterized by
\be
r \cos \theta' - \sqrt{r^2 - R^2} \cosh(t'/R) > 0\,.
\ee

A final useful bit of geometry is the antipodal map on AdS, which acts by changing the sign of
the embedding coordinates.
\be
A \, : \, (u,v,{\bf x},y) \rightarrow (-u,-v,-{\bf x},-y)
\ee
In Rindler coordinates this is realized by
\be
\label{antipodal}
A \, : \, (t,r,\phi,{\bf n}) \rightarrow (t + i \pi R,r,\phi + i \pi,{\bf n})
\ee
Under the antipodal map $\sigma(x \vert Ay) = - \sigma(x \vert y)$, and for a field of integer dimension $\phi(Ax) = (-1)^\Delta \phi(x)$.
This lets us write the smearing function for a point inside the horizon.  Assuming $\Delta$ is an integer
\bea
\label{InsideRindlerSmear}
&& \phi(t=\phi=0,r < R) \\
\nonumber
&& = c_{d\Delta}
\int_{\sigma > 0} dt' d\theta' d\Omega'_{d-2} R^{d-1} \sin^{d-2} \theta' \left(r \cos \theta' + \sqrt{R^2 - r^2} \sinh {t' \over R}\right)^{\Delta-d}
{\cal O}_R(t',i\theta',{\bf n}') \\
\nonumber
&& + (-1)^\Delta c_{d\Delta}
\int_{\sigma < 0} dt' d\theta' d\Omega'_{d-2} R^{d-1} \sin^{d-2} \theta' \left(-r \cos \theta' + \sqrt{R^2 - r^2} \sinh {t' \over R}\right)^{\Delta-d}
{\cal O}_L(t',i\theta',{\bf n}')
\eea
The generalization to non-integer $\Delta$ can be found in \cite{Hamilton:2006fh}.

\section{CFT correlators at finite temperature\label{appendix:CFT}}

In Minkowski space the 2-point correlator for operators of dimension $\Delta$ is fixed by conformal
invariance.
\be
\label{Minkowski2point}
\langle {\cal O}(T,{\bf X}) {\cal O}(0,0) \rangle = {1 \over \left(-T^2 + \vert {\bf X} \vert^2 \right)^\Delta}
\ee
This is the correlator one would use in Poincar\'e coordinates, where the boundary metric
is $ds^2_{\rm Poincare \,\, bdy} = - dT^2 + \vert d{\bf X} \vert^2$.  Changing coordinates on the boundary using (\ref{t}), (\ref{phi}), one finds that\footnote{There's no real need to do
the change of coordinates.  One can read this off by comparing (\ref{RindlerMetric}) and (\ref{PoincareMetric}).}
\be
ds^2_{\rm Poincare \,\, bdy} = \lim_{Z \rightarrow 0} {Z^2 r^2 \over R^4} \left(-dt^2 + R^2 ds^2_{{\cal H}^{d-1}}\right)
\ee
Dropping the conformal factor, we identify the quantity in parenthesis with the Rindler boundary metric
$ds^2_{\rm Rindler \,\, bdy} = -dt^2 + R^2 ds^2_{{\cal H}^{d-1}}$.  In Rindler coordinates the correlator (\ref{Minkowski2point}) becomes\footnote{The easiest way to see this is to note that the CFT correlator is the boundary limit of
$\left(2 Z Z' \sigma\right)^{-\Delta}$ in Poincar\'e coordinates, or the boundary limit of $\left(2 R^4 \sigma / r r'\right)^{-\Delta}$
in Rindler coordinates.}
\be
\label{Rindler2point}
\langle {\cal O}(t,\phi,{\bf n}) {\cal O}(t',\phi',{\bf n}') \rangle = (2 R^2)^{-\Delta} \left(\cosh \phi \cosh \phi' -
\sinh \phi \sinh \phi' \, {\bf n} \cdot {\bf n}' - \cosh {t - t' \over R} \right)^{-\Delta}
\ee
This is now a thermal correlator, periodic in imaginary time with period $2 \pi R$.  The late-time behavior
of the correlator is
\be
\langle {\cal O}(t) {\cal O}(0) \rangle \sim e^{-\Delta t / R} = e^{-2\pi\Delta t / \beta}
\ee
as claimed in (\ref{ExponentialDecay}).  Note that this behavior is fixed by conformal invariance.

The result (\ref{Rindler2point}) is appropriate for two operators on the same boundary in the thermofield double formalism.  An operator on the
on the left boundary can be obtained by shifting $t \rightarrow t + i \pi R$, so the left - right correlator is
\be
\label{Rindler2pointLR}
\langle {\cal O}_L(t,\phi,{\bf n}) {\cal O}_R(t',\phi',{\bf n}') \rangle = (2 R^2)^{-\Delta} \left(\cosh \phi \cosh \phi' -
\sinh \phi \sinh \phi' \, {\bf n} \cdot {\bf n}' + \cosh {t - t' \over R} \right)^{-\Delta}
\ee

\section{Correlators at finite entropy\label{appendix:timescales}}

In appendix \ref{appendix:CFT} we studied the time dependence of a CFT correlator in the thermodynamic limit and found a universal
exponential decay fixed by conformal invariance.  Here we are interested in the behavior of correlators at finite entropy.  We consider
a correlator $C(t) = \langle \psi \vert {\cal O}(t) {\cal O}(0) \vert \psi \rangle$ in a typical pure state of the system and ask for
the probability distribution which governs the different possible values of the correlator.

The exact distribution depends on the matrix elements of the operator ${\cal O}$.  But we are mostly interested in how the distribution
depends on the dimension of the available Hilbert space, so we will model the correlator as the inner product of two unit vectors
$C = \langle \psi_1 \vert \psi_2 \rangle$.  For a generic Hamiltonian we expect that $\vert \psi_1 \rangle$ and $\vert \psi_2 \rangle$
should be chosen randomly.  In the Hilbert space ${\cal H} = {\mathbb C}^N$ we take
$\vert \psi_1 \rangle = (1,0,\ldots,0)$ without loss of generality, and we take $\vert \psi_2 \rangle$ to be chosen at random on the unit sphere $S^{2N-1} \subset {\mathbb C}^N$.
We write the metric on this sphere
\be
d\Omega^2_{S^{2N-1}} = d\theta^2 + \sin^2 \theta d\phi^2 + \sin^2 \theta \sin^2 \phi \, d\Omega^2_{S^{2N-3}}
\ee
with $0 \leq \theta,\phi \leq \pi$ and embed the sphere in ${\mathbb C}^N$ by setting
\be
\vert \psi_2 \rangle = (\cos \theta + i \sin \theta \cos \phi,\ldots)
\ee
Geometrically we've represented $S^{2N-1}$ as a bundle over the unit disc with fiber $S^{2N-3}$.  (The fibers degenerate at the edge of the disc.)
The correlator is then modeled by $C = \langle \psi_1 \vert \psi_2 \rangle = \cos \theta + i \sin \theta \cos \phi$.

With $\vert \psi_2 \rangle$ distributed uniformly according to the volume form on $S^{2N-1}$ we can integrate over $S^{2N-3}$ to get
the differential probability for having a given inner product.
\be
dP = {{\rm vol}\left(S^{2N-3}\right) \over {\rm vol}\left(S^{2N-1}\right)} \, d\theta \sin \theta d\phi \left(\sin \theta \sin \phi\right)^{2N-3}
\ee
In terms of $C$ the probability is
\be
dP = {N - 1 \over \pi} \left(1 - \Vert C \Vert^2 \right)^{N-2} \, d({\rm Re} \, C) d({\rm Im} \, C)
\ee
For large $N$
\be
\label{Gaussian}
dP \approx {N \over \pi} \, \exp\left({-N \Vert C \Vert^2}\right) \, d({\rm Re} \, C) d({\rm Im} \, C)
\ee
So the correlator obeys a Gaussian distribution with variance $\langle \Vert C \Vert^2 \rangle = {1 \over N}$.  Note that (\ref{Gaussian})
is valid for $N \Vert C \Vert^4 \ll 1$; since $\log (1 - x^2) < -x^2$ the true distribution falls faster than Gaussian.

The picture that results from modeling a correlator as a generic inner product of two unit vectors is that most of the time correlators undergo
random fluctuations of size $1/\sqrt{N} = e^{-S/2}$.  Fluctuations of ${\cal O}(1)$ happen with probability $\sim e^{-N}$ and therefore occur
on timescales of order $e^{N} = \exp\left(e^S\right)$.  This phenomenon has been studied in more detail in \cite{Barbon:2014rma}.


\begin{thebibliography}{10}

\bibitem{'tHooft:1993gx}
G.~'t~Hooft, ``{Dimensional reduction in quantum gravity},''
\href{http://arxiv.org/abs/gr-qc/9310026}{{\ttfamily arXiv:gr-qc/9310026
  [gr-qc]}}.

\bibitem{Susskind:1994vu}
L.~Susskind, ``{The world as a hologram},''
  \href{http://dx.doi.org/10.1063/1.531249}{{\em J.Math.Phys.} {\bfseries 36}
  (1995) 6377--6396},
\href{http://arxiv.org/abs/hep-th/9409089}{{\ttfamily arXiv:hep-th/9409089
  [hep-th]}}.

\bibitem{Balasubramanian:1998sn}
V.~Balasubramanian, P.~Kraus, and A.~E. Lawrence, ``{Bulk vs. boundary dynamics
  in anti-de Sitter spacetime},''
  \href{http://dx.doi.org/10.1103/PhysRevD.59.046003}{{\em Phys. Rev.}
  {\bfseries D59} (1999) 046003},
\href{http://arxiv.org/abs/hep-th/9805171}{{\ttfamily arXiv:hep-th/9805171}}.

\bibitem{Banks:1998dd}
T.~Banks, M.~R. Douglas, G.~T. Horowitz, and E.~J. Martinec, ``{AdS dynamics
  from conformal field theory},''
\href{http://arxiv.org/abs/hep-th/9808016}{{\ttfamily arXiv:hep-th/9808016}}.

\bibitem{Dobrev:1998md}
V.~K. Dobrev, ``{Intertwining operator realization of the AdS/CFT
  correspondence},''
  \href{http://dx.doi.org/10.1016/S0550-3213(99)00284-9}{{\em Nucl. Phys.}
  {\bfseries B553} (1999) 559--582},
\href{http://arxiv.org/abs/hep-th/9812194}{{\ttfamily arXiv:hep-th/9812194}}.

\bibitem{Bena:1999jv}
I.~Bena, ``{On the construction of local fields in the bulk of AdS(5) and other
  spaces},'' \href{http://dx.doi.org/10.1103/PhysRevD.62.066007}{{\em Phys.
  Rev.} {\bfseries D62} (2000) 066007},
\href{http://arxiv.org/abs/hep-th/9905186}{{\ttfamily arXiv:hep-th/9905186}}.

\bibitem{Hamilton:2005ju}
A.~Hamilton, D.~Kabat, G.~Lifschytz, and D.~A. Lowe, ``Local bulk operators in
  AdS/CFT: A boundary view of horizons and locality,'' {\em Phys. Rev.}
  {\bfseries D73} (2006) 086003,
\href{http://arxiv.org/abs/hep-th/0506118}{{\ttfamily hep-th/0506118}}.

\bibitem{Hamilton:2006az}
A.~Hamilton, D.~Kabat, G.~Lifschytz, and D.~A. Lowe, ``Holographic
  representation of local bulk operators,'' {\em Phys. Rev.} {\bfseries D74}
  (2006) 066009,
\href{http://arxiv.org/abs/hep-th/0606141}{{\ttfamily hep-th/0606141}}.

\bibitem{Hamilton:2006fh}
A.~Hamilton, D.~Kabat, G.~Lifschytz, and D.~A. Lowe, ``Local bulk operators in
  AdS/CFT: A holographic description of the black hole interior,'' {\em Phys.
  Rev.} {\bfseries D75} (2007) 106001,
\href{http://arxiv.org/abs/hep-th/0612053}{{\ttfamily hep-th/0612053}}.

\bibitem{Heemskerk:2012mq}
I.~Heemskerk, ``{Construction of bulk fields with gauge redundancy},''
\href{http://arxiv.org/abs/1201.3666}{{\ttfamily arXiv:1201.3666 [hep-th]}}.

\bibitem{Kabat:2012hp}
D.~Kabat, G.~Lifschytz, S.~Roy, and D.~Sarkar, ``{Holographic representation of
  bulk fields with spin in AdS/CFT},''
  \href{http://dx.doi.org/10.1103/PhysRevD.86.026004,
  10.1103/PhysRevD.86.029901}{{\em Phys.Rev.} {\bfseries D86} (2012) 026004},
\href{http://arxiv.org/abs/1204.0126}{{\ttfamily arXiv:1204.0126 [hep-th]}}.

\bibitem{Kabat:2011rz}
D.~Kabat, G.~Lifschytz, and D.~A. Lowe, ``{Constructing local bulk observables
  in interacting AdS/CFT},''
  \href{http://dx.doi.org/10.1103/PhysRevD.83.106009}{{\em Phys.Rev.}
  {\bfseries D83} (2011) 106009},
\href{http://arxiv.org/abs/1102.2910}{{\ttfamily arXiv:1102.2910 [hep-th]}}.

\bibitem{Heemskerk:2012mn}
I.~Heemskerk, D.~Marolf, and J.~Polchinski, ``{Bulk and transhorizon
  measurements in AdS/CFT},''
\href{http://arxiv.org/abs/1201.3664}{{\ttfamily arXiv:1201.3664 [hep-th]}}.

\bibitem{Kabat:2012av}
D.~Kabat and G.~Lifschytz, ``{CFT representation of interacting bulk gauge
  fields in AdS},'' \href{http://dx.doi.org/10.1103/PhysRevD.87.086004}{{\em
  Phys.Rev.} {\bfseries D87} (2013) 086004},
\href{http://arxiv.org/abs/1212.3788}{{\ttfamily arXiv:1212.3788 [hep-th]}}.

\bibitem{Kabat:2013wga}
D.~Kabat and G.~Lifschytz, ``{Decoding the hologram: Scalar fields interacting
  with gravity},''
\href{http://arxiv.org/abs/1311.3020}{{\ttfamily arXiv:1311.3020 [hep-th]}}.

\bibitem{Bousso:1999xy}
R.~Bousso, ``{A covariant entropy conjecture},''
  \href{http://dx.doi.org/10.1088/1126-6708/1999/07/004}{{\em JHEP} {\bfseries
  9907} (1999) 004},
\href{http://arxiv.org/abs/hep-th/9905177}{{\ttfamily arXiv:hep-th/9905177
  [hep-th]}}.

\bibitem{Susskind:1998dq}
L.~Susskind and E.~Witten, ``{The holographic bound in anti-de Sitter space},''
\href{http://arxiv.org/abs/hep-th/9805114}{{\ttfamily arXiv:hep-th/9805114
  [hep-th]}}.

\bibitem{Mathur:2009hf}
S.~D. Mathur, ``{The information paradox: A pedagogical introduction},''
  \href{http://dx.doi.org/10.1088/0264-9381/26/22/224001}{{\em
  Class.Quant.Grav.} {\bfseries 26} (2009) 224001},
\href{http://arxiv.org/abs/0909.1038}{{\ttfamily arXiv:0909.1038 [hep-th]}}.

\bibitem{Mathur:2005zp}
S.~D. Mathur, ``{The fuzzball proposal for black holes: An elementary
  review},'' \href{http://dx.doi.org/10.1002/prop.200410203}{{\em
  Fortsch.Phys.} {\bfseries 53} (2005) 793--827},
\href{http://arxiv.org/abs/hep-th/0502050}{{\ttfamily arXiv:hep-th/0502050
  [hep-th]}}.

\bibitem{Mathur:2008nj}
S.~D. Mathur, ``{Fuzzballs and the information paradox: A summary and
  conjectures},''
\href{http://arxiv.org/abs/0810.4525}{{\ttfamily arXiv:0810.4525 [hep-th]}}.

\bibitem{Almheiri:2012rt}
A.~Almheiri, D.~Marolf, J.~Polchinski, and J.~Sully, ``{Black holes:
  Complementarity or firewalls?},''
  \href{http://dx.doi.org/10.1007/JHEP02(2013)062}{{\em JHEP} {\bfseries 1302}
  (2013) 062},
\href{http://arxiv.org/abs/1207.3123}{{\ttfamily arXiv:1207.3123 [hep-th]}};
cf.\ S.~L.~Braunstein, ``{Black hole entropy as entropy of
entanglement, or it's curtains for the equivalence principle},''
\href{http://arxiv.org/abs/0907.1190v1}{{\ttfamily arXiv:0907.1190v1 [quant-ph]}}.

\bibitem{Almheiri:2013hfa}
A.~Almheiri, D.~Marolf, J.~Polchinski, D.~Stanford, and J.~Sully, ``{An
  apologia for firewalls},''
  \href{http://dx.doi.org/10.1007/JHEP09(2013)018}{{\em JHEP} {\bfseries 1309}
  (2013) 018},
\href{http://arxiv.org/abs/1304.6483}{{\ttfamily arXiv:1304.6483 [hep-th]}}.

\bibitem{Giddings:2011ks}
S.~B. Giddings, ``{Models for unitary black hole disintegration},''
  \href{http://dx.doi.org/10.1103/PhysRevD.85.044038}{{\em Phys.Rev.}
  {\bfseries D85} (2012) 044038},
\href{http://arxiv.org/abs/1108.2015}{{\ttfamily arXiv:1108.2015 [hep-th]}}.

\bibitem{Giddings:2012gc}
S.~B. Giddings, ``{Nonviolent nonlocality},''
  \href{http://dx.doi.org/10.1103/PhysRevD.88.064023}{{\em Phys.Rev.}
  {\bfseries D88} (2013) 064023},
\href{http://arxiv.org/abs/1211.7070}{{\ttfamily arXiv:1211.7070 [hep-th]}}.

\bibitem{Giddings:2013noa}
S.~B. Giddings and Y.~Shi, ``{Effective field theory models for nonviolent
  information transfer from black holes},''
\href{http://arxiv.org/abs/1310.5700}{{\ttfamily arXiv:1310.5700 [hep-th]}}.

\bibitem{Jevicki:1998rr}
A.~Jevicki and S.~Ramgoolam, ``{Noncommutative gravity from the AdS / CFT
  correspondence},''
  \href{http://dx.doi.org/10.1088/1126-6708/1999/04/032}{{\em JHEP} {\bfseries
  9904} (1999) 032},
\href{http://arxiv.org/abs/hep-th/9902059}{{\ttfamily arXiv:hep-th/9902059
  [hep-th]}}.

\bibitem{Garner:2014kna}
D.~Garner, S.~Ramgoolam, and C.~Wen, ``{Thresholds of large N factorization in
  CFT4 : Exploring bulk locality in AdS5},''
\href{http://arxiv.org/abs/1403.5281}{{\ttfamily arXiv:1403.5281 [hep-th]}}.

\bibitem{Hamilton:2007wj}
A.~Hamilton, D.~N. Kabat, G.~Lifschytz, and D.~A. Lowe, ``{Local bulk operators
  in AdS/CFT and the fate of the BTZ singularity},''
\href{http://arxiv.org/abs/0710.4334}{{\ttfamily arXiv:0710.4334 [hep-th]}}.

\bibitem{Lowe:2009mq}
D.~A. Lowe, ``{Black hole complementarity from AdS/CFT},''
  \href{http://dx.doi.org/10.1103/PhysRevD.79.106008}{{\em Phys. Rev.}
  {\bfseries D79} (2009) 106008},
\href{http://arxiv.org/abs/0903.1063}{{\ttfamily arXiv:0903.1063 [hep-th]}}.

\bibitem{Rey:2014dpa}
S.-J. Rey and V.~Rosenhaus, ``{Scanning tunneling macroscopy, black holes, and
  AdS/CFT bulk locality},''
\href{http://arxiv.org/abs/1403.3943}{{\ttfamily arXiv:1403.3943 [hep-th]}}.

\bibitem{Morrison:2014jha}
I.~A. Morrison, ``{Boundary-to-bulk maps for AdS causal wedges and the
  Reeh-Schlieder property in holography},''
\href{http://arxiv.org/abs/1403.3426}{{\ttfamily arXiv:1403.3426 [hep-th]}}.

\bibitem{Papadodimas:2012aq}
K.~Papadodimas and S.~Raju, ``{An infalling observer in AdS/CFT},''
  \href{http://dx.doi.org/10.1007/JHEP10(2013)212}{{\em JHEP} {\bfseries 1310}
  (2013) 212},
\href{http://arxiv.org/abs/1211.6767}{{\ttfamily arXiv:1211.6767 [hep-th]}}.

\bibitem{Susskind:2014rva}
L.~Susskind, ``{Computational complexity and black hole horizons},''
\href{http://arxiv.org/abs/1402.5674}{{\ttfamily arXiv:1402.5674 [hep-th]}}.

\bibitem{Maldacena:2001kr}
J.~M. Maldacena, ``Eternal black holes in anti-de sitter,'' {\em JHEP}
  {\bfseries 04} (2003) 021,
\href{http://arxiv.org/abs/hep-th/0106112}{{\ttfamily hep-th/0106112}}.

\bibitem{Dyson:2002nt}
L.~Dyson, J.~Lindesay, and L.~Susskind, ``{Is there really a de Sitter/CFT
  duality?},'' \href{http://dx.doi.org/10.1088/1126-6708/2002/08/045}{{\em
  JHEP} {\bfseries 0208} (2002) 045},
\href{http://arxiv.org/abs/hep-th/0202163}{{\ttfamily arXiv:hep-th/0202163
  [hep-th]}}.

\bibitem{Barbon:2003aq}
J.~L.~F. Barbon and E.~Rabinovici, ``{Very long time scales and black hole
  thermal equilibrium},'' {\em JHEP} {\bfseries 11} (2003) 047,
\href{http://arxiv.org/abs/hep-th/0308063}{{\ttfamily arXiv:hep-th/0308063}}.

\bibitem{Barbon:2014rma}
J.~L.~F. Barbon and E.~Rabinovici, ``{Geometry and quantum noise},''
\href{http://arxiv.org/abs/1404.7085}{{\ttfamily arXiv:1404.7085 [hep-th]}}.

\bibitem{Marolf:2013dba}
D.~Marolf and J.~Polchinski, ``{Gauge/gravity duality and the black hole
  interior},'' \href{http://dx.doi.org/10.1103/PhysRevLett.111.171301}{{\em
  Phys.Rev.Lett.} {\bfseries 111} (2013) 171301},
\href{http://arxiv.org/abs/1307.4706}{{\ttfamily arXiv:1307.4706 [hep-th]}}.

\bibitem{Mathur:2011wg}
S.~D. Mathur and C.~J. Plumberg, ``{Correlations in Hawking radiation and the
  infall problem},'' \href{http://dx.doi.org/10.1007/JHEP09(2011)093}{{\em
  JHEP} {\bfseries 1109} (2011) 093},
\href{http://arxiv.org/abs/1101.4899}{{\ttfamily arXiv:1101.4899 [hep-th]}}.

\bibitem{Birmingham:2002ph}
D.~Birmingham, I.~Sachs, and S.~N. Solodukhin, ``{Relaxation in conformal field
  theory, Hawking-Page transition, and quasinormal/normal modes},''
  \href{http://dx.doi.org/10.1103/PhysRevD.67.104026}{{\em Phys. Rev.}
  {\bfseries D67} (2003) 104026},
\href{http://arxiv.org/abs/hep-th/0212308}{{\ttfamily arXiv:hep-th/0212308}}.

\bibitem{Mann:1997iz}
R.~B. Mann, ``{Topological black holes: Outside looking in},''
\href{http://arxiv.org/abs/gr-qc/9709039}{{\ttfamily arXiv:gr-qc/9709039
  [gr-qc]}}.

\bibitem{Birmingham:1998nr}
D.~Birmingham, ``{Topological black holes in Anti-de Sitter space},''
  \href{http://dx.doi.org/10.1088/0264-9381/16/4/009}{{\em Class.Quant.Grav.}
  {\bfseries 16} (1999) 1197--1205},
\href{http://arxiv.org/abs/hep-th/9808032}{{\ttfamily arXiv:hep-th/9808032
  [hep-th]}}.

\bibitem{Ichinose:1994rg}
I.~Ichinose and Y.~Satoh, ``{Entropies of scalar fields on three-dimensional
  black holes},'' \href{http://dx.doi.org/10.1016/0550-3213(95)00197-Z}{{\em
  Nucl.Phys.} {\bfseries B447} (1995) 340--372},
\href{http://arxiv.org/abs/hep-th/9412144}{{\ttfamily arXiv:hep-th/9412144
  [hep-th]}}.

\bibitem{Lifschytz:1993eb}
G.~Lifschytz and M.~Ortiz, ``{Scalar field quantization on the
  (2+1)-dimensional black hole background},''
  \href{http://dx.doi.org/10.1103/PhysRevD.49.1929}{{\em Phys.Rev.} {\bfseries
  D49} (1994) 1929--1943},
\href{http://arxiv.org/abs/gr-qc/9310008}{{\ttfamily arXiv:gr-qc/9310008
  [gr-qc]}}.

\bibitem{Papadodimas:2013jku}
K.~Papadodimas and S.~Raju, ``{State-dependent bulk-boundary maps and black
  hole complementarity},''
\href{http://arxiv.org/abs/1310.6335}{{\ttfamily arXiv:1310.6335 [hep-th]}}.

\bibitem{Hawking:1976ra}
S.~W. Hawking, ``{Breakdown of predictability in gravitational collapse},''
\href{http://dx.doi.org/10.1103/PhysRevD.14.2460}{{\em Phys. Rev.} {\bfseries
  D14} (1976) 2460--2473}.

\bibitem{Lowe:2006xm}
D.~A. Lowe and L.~Thorlacius, ``{Comments on the black hole information
  problem},'' \href{http://dx.doi.org/10.1103/PhysRevD.73.104027}{{\em
  Phys.Rev.} {\bfseries D73} (2006) 104027},
\href{http://arxiv.org/abs/hep-th/0601059}{{\ttfamily arXiv:hep-th/0601059
  [hep-th]}}.

\bibitem{Papadodimas:2013kwa}
K.~Papadodimas and S.~Raju, ``{The unreasonable effectiveness of exponentially
  suppressed corrections in preserving information},''
\href{http://dx.doi.org/10.1142/S0218271813420303}{{\em Int.J.Mod.Phys.}
  {\bfseries D22} (2013) 1342030}.

\bibitem{Page:1993df}
D.~N. Page, ``{Expected entropy of a subsystem},''
  \href{http://dx.doi.org/10.1103/PhysRevLett.71.1291}{{\em Phys. Rev. Lett.}
  {\bfseries 71} (1993) 1291--1294},
\href{http://arxiv.org/abs/gr-qc/9305007}{{\ttfamily arXiv:gr-qc/9305007}}.

\bibitem{Page:1993wv}
D.~N. Page, ``{Information in black hole radiation},''
  \href{http://dx.doi.org/10.1103/PhysRevLett.71.3743}{{\em Phys. Rev. Lett.}
  {\bfseries 71} (1993) 3743--3746},
\href{http://arxiv.org/abs/hep-th/9306083}{{\ttfamily arXiv:hep-th/9306083}}.

\bibitem{Susskind:1993if}
L.~Susskind, L.~Thorlacius, and J.~Uglum, ``{The stretched horizon and black
  hole complementarity},''
  \href{http://dx.doi.org/10.1103/PhysRevD.48.3743}{{\em Phys. Rev.} {\bfseries
  D48} (1993) 3743--3761},
\href{http://arxiv.org/abs/hep-th/9306069}{{\ttfamily arXiv:hep-th/9306069}}.

\bibitem{Susskind:1993mu}
L.~Susskind and L.~Thorlacius, ``{Gedanken experiments involving black
  holes},'' \href{http://dx.doi.org/10.1103/PhysRevD.49.966}{{\em Phys. Rev.}
  {\bfseries D49} (1994) 966--974},
\href{http://arxiv.org/abs/hep-th/9308100}{{\ttfamily arXiv:hep-th/9308100}}.

\bibitem{Iizuka:2013ria}
N.~Iizuka and D.~Kabat, ``{On the mutual information in Hawking radiation},''
  \href{http://dx.doi.org/10.1103/PhysRevD.88.084010}{{\em Phys.Rev.}
  {\bfseries D88} (2013) 084010},
\href{http://arxiv.org/abs/1308.2386}{{\ttfamily arXiv:1308.2386 [hep-th]}}.

\bibitem{Hawking:1982dh}
S.~W. Hawking and D.~N. Page, ``{Thermodynamics of black holes in anti-de
  Sitter space},''
\href{http://dx.doi.org/10.1007/BF01208266}{{\em Commun. Math. Phys.}
  {\bfseries 87} (1983) 577}.

\bibitem{Witten:1998zw}
E.~Witten, ``{Anti-de Sitter space, thermal phase transition, and confinement
  in gauge theories},'' {\em Adv.Theor.Math.Phys.} {\bfseries 2} (1998)
  505--532,
\href{http://arxiv.org/abs/hep-th/9803131}{{\ttfamily arXiv:hep-th/9803131
  [hep-th]}}.

\bibitem{Parikh:2012kg}
M.~Parikh and P.~Samantray, ``{Rindler-AdS/CFT},''
\href{http://arxiv.org/abs/1211.7370}{{\ttfamily arXiv:1211.7370 [hep-th]}}.

\end{thebibliography}

\providecommand{\href}[2]{#2}\begingroup\raggedright\endgroup

\end{document}